\documentclass[12pt]{article}
\usepackage{cmap} %
\usepackage[paper=a4paper,left=1.91cm,right=1.91cm, top=2.54cm, bottom=2.54cm]{geometry}
\usepackage{amsmath, amsfonts, amssymb, amsthm, array, graphicx, float, morefloats, setspace, wrapfig}
\usepackage{icomma}
\usepackage{kpfonts}
\usepackage[T1]{fontenc}
\usepackage[utf8]{inputenc}
\usepackage[english]{babel}
\usepackage[font=small]{caption}

\usepackage[title]{appendix}
\usepackage{appendix}

\usepackage[protrusion=true,expansion=all]{microtype}
\LoadMicrotypeFile{cmr}
\SetProtrusion
{ encoding = *,
 family = * }
{
« = {1000,     },
» = {    , 1000},
„ = {1000,     },
“ = {    , 1000},
” = {    , 1000},
( = {1000,     },
) = {    , 1000},
! = {    , 1000},
? = {    , 1000},
: = {    , 1000},
; = {    , 1000},
. = {    , 1000},
- = {    ,  500},
— = {    ,  500},
{,} = {    , 1000}
}
\theoremstyle{definition}

\usepackage{rotating} 

\usepackage{booktabs}
\usepackage{tablefootnote}
\usepackage{threeparttable}

\usepackage{color}

\usepackage[section]{placeins}
\usepackage{float}

\usepackage{tikz}

\usepackage{verbatim}

\usepackage[labelfont=bf,skip=0pt]{caption}
\newcommand{\mycaption}[2]{\caption[#1]{\textbf{#1.} #2}}
\usepackage[labelformat=simple,labelfont=rm]{subcaption}

\newcommand{\argmin}{\operatornamewithlimits{argmin}}
\newcommand{\argmax}{\operatornamewithlimits{argmax}}

\usepackage{bm}
\usepackage{tcolorbox} %

\usepackage{dsfont}
 
\newtheorem{theorem}{Theorem}
\newtheorem{example}{Example}
\newtheorem{lemma}{Lemma}
\newtheorem{assumption}{Assumption}
\newtheorem{remark}{Remark}

\usepackage[backend=biber, bibencoding=utf8, sorting=nty, style=chicago-authordate, maxcitenames=2, maxbibnames=99, style=bwl-FU, uniquename=false, uniquelist=false, natbib=true]{biblatex}
\renewbibmacro{in:}{\ifentrytype{article}{}{\printtext{\bibstring{in}\intitlepunct}}}
\DeclareFieldFormat[article,inbook,incollection,inproceedings,patent,thesis,unpublished]{citetitle}{#1}
\DeclareFieldFormat[article,inbook,incollection,inproceedings,patent,thesis,unpublished]{title}{#1}

\usepackage[autostyle=true]{csquotes}

\makeatletter
\def\@fnsymbol#1{\ensuremath{\ifcase#1\or \dagger\or \ddagger\or
\mathsection\or \mathparagraph\or \|\or **\or \dagger\dagger
\or \ddagger\ddagger \else\@ctrerr\fi}}
\makeatother

\usepackage[affil-it]{authblk}

\title{Frequentist Shrinkage Under Inequality Constraints}
\author{Edvard Bakhitov\thanks{I am grateful to Xu Cheng and Frank DiTraglia for their support and encouragement. Special thanks to Max Kasy for providing the dataset. This paper benefitted from feedback from Karun Adusumilli,  St{\'e}phane Bonhomme, Philippe Goulet Coulombe, Phillip Heiler, Toru Kitagawa, and Frank Schorfheide, as well as seminar participants at the UPenn econometrics lunch, and participants at the ESWM 2019 conference in Rotterdam.}}
\affil{University of Pennsylvania}
\date{\today}

\begin{document}

\maketitle

\begin{abstract}
   This paper shows how to shrink extremum estimators towards inequality constraints motivated by economic theory. We propose an Inequality Constrained Shrinkage Estimator (ICSE) which takes the form of a weighted average between the unconstrained and inequality constrained estimators with the data dependent weight. The weight drives both the direction and degree of shrinkage. We use a local asymptotic framework to derive the asymptotic distribution and risk of the ICSE. We provide conditions under which the asymptotic risk of the ICSE is strictly less than that of the unrestricted extremum estimator. The degree of shrinkage cannot be consistently estimated under the local asymptotic framework. To address this issue, we propose a feasible plug-in estimator and investigate its finite sample behavior. We also apply our framework to gasoline demand estimation under the Slutsky restriction. 
   
\end{abstract}


\section{Introduction}

Inequality constraints are common in applied economic research. Typical examples are monotonicity constraints on utility or production functions, restrictions on estimated covariance matrices such as positive definiteness, restrictions on the Slutsky matrix, etc. If the imposed constraints hold, we will get more efficient estimates. If not, the estimates will be biased. 

The paper proposes an alternative way to use economic theory in estimation. We introduce a generalized shrinkage estimator that shrinks an estimator that ignores theoretical restrictions towards inequality constraints motivated by theory. The inequality constrained shrinkage estimator (ICSE) takes a simple weighted average form between the unconstrained and inequality constrained estimators, with the data-driven weight inversely proportional to the loss function evaluated at the two estimates. We show that the degree of shrinkage depends on which constraints bind, thus, both the direction and degree of shrinkage are fully data driven. 

We show that under certain conditions the ICSE outperforms the unrestricted estimator, regardless of what the true data generating process is and whether the theory is correct or not. We demonstrate that the ICSE has a smaller asymptotic risk than the unrestricted estimator uniformly over the parameter space local to the restricted (shrinkage) parameter space. The theory we present applies to a large set of extremum estimators, such as the Generalized Method of Moments (GMM) estimator, the Maximum Likelihood estimator (MLE), the Minimum Distance (MD) estimator, etc.

We use the local asymptotic framework to analyze the performance of the ICSE. To be precise, we assume that the parameter space is located in a $n^{-1/2}$-neighborhood of the restricted space, reflecting the belief that the imposed theoretical restrictions are only ''approximately correct''. In contrast to the generalized James-Stein estimator, the asymptotic distribution of the ICSE is not normal. Since the ICSE is a weighted average of the unconstrained and inequality constrained estimators, the asymptotic distribution of the shrinkage estimator inherits the non-normality of the inequality constrained estimator. 

Under the local asymptotic framework, it is impossible to consistently estimate the optimal degree of shrinkage, as it depends on the local $O(n^{-1/2})$ parameters (see e.g. \citealp{hjort_claeskens2003}). To address this issue, we propose a feasible plug-in estimator based on the asymptotically unbiased estimator of the local parameters. However, this makes the estimated shrinkage parameter asymptotically random, which affects the asymptotic distribution of the averaging weight. As a result, the feasible estimator is not consistent and the dominance result may not hold. 

In our Monte Carlo study we investigate the finite sample performance of the feasible ICSE along with the generalized James-Stein estimator of \cite{hansen2016}, the Empirical Bayes (EB) estimator, the unrestricted estimator, and the restricted estimator. Simulations show that the feasible ICSE dominates the unrestricted estimator in terms of mean squared error. Moreover, it also dominates the generalized James-Stein estimator in cases when a subset of the constraints bind. We also show that the ICSE performs better than the EB estimator when the constraints are violated or close to bind, while the EB estimator dominates the ICSE when the constraints are satisfied as strict inequalities.

In our application we consider gasoline demand estimation under the Slutsky restriction. We estimate the demand curves across three income groups corresponding to the first, second, and third quartiles, respectively. We show that the shrinkage effect is more prominent for the low income group, since consumers with low income are less likely to have upward sloping demand curves. In a similar application, \cite{kasy2018} use the Empirical Bayes framework to show that the degree of shrinkage is similar across different groups.

\vspace{1em}

The literature on shrinkage estimation begins with \cite{stein1956} who observed that the unconstrained estimator in a Gaussian location model is inadmissible when the dimension of the parameter vector is greater than two. This lead to a seminal paper by \cite{james_stein1961} where they proposed a shrinkage estimator that dominates the MLE. \cite{baranchik1964} showed that the James-Stein estimator is inadmissible and dominated by its positive part version. However, even the positive part James-Stein estimator is inadmissible. \cite{shao1994} propose a piecewise linear estimator that has even smaller risk. Theory for risk analysis of shrinkage estimators was provided by \cite{stein1981}. \cite{hansen2015} compares the performance of different shrinkage estimators and provides corresponding efficiency bounds.

All of the aforementioned estimators shrink the parameters towards zero. In contrast, \cite{oman1982a,oman1982b} introduce estimators which shrink towards linear subspaces. \cite{delNegro2004} show how to shrink to non-linear subspaces in the Bayesian framework, using a DSGE model-based prior to estimate the VAR impulse response functions. In their recent paper, \cite{kasy2018} provide an Empirical Bayes framework which allows to shrink to various theoretical restrictions in form of both equalities and inequalities. Our paper complements the aforementioned literature by extending the Stein's type shrinkage argument to non-linear inequality constraints.

\cite{james_stein1961} first showed that the shrinkage estimator dominates the unrestricted MLE in exact normal sampling. \cite{hansen2016} provides a generalized James-Stein type estimator for parametric models and shows that it dominates the MLE in a pointwise locally asymptotic sense.\footnote{For a given real vector $c$, the pointwise local asymptotic analysis considers a sequence of localized parameters $\theta_{n} = cn^{-1/2}$, and  derives  the  asymptotic  (truncated)  risk  of  the  averaging  estimator under $\theta_{n}$ for given $c$. Such analysis will produce a pointwise risk function for the shrinkage estimator.} \cite{hansen2017} shows that a shrinkage estimator that shrinks the OLS estimator towards the 2SLS estimator has a smaller asymptotic risk than the ordinary OLS estimator. \cite{ditraglia2016} studies the averaging GMM estimator with the averaging weight based on the focused moment selection criterion. The results in the paper suggest that the averaging estimator does not uniformly dominate the conservative estimator. Unlike the aforementioned papers using the pointwise local asymptotic framework, \cite{cheng_et_al2019} establish the uniform dominance result of the GMM averaging estimator over the conservative estimator.

This paper is also closely related to the frequentist model averaging literature. \cite{hansen2007} introduces a model averaging estimator for linear nested models and shows that it is asymptotically optimal. He proposes to minimize a Mallows criterion to select the model weights, which is asymptotically equivalent to minimizing the squared error. \cite{wan_et_al2010} show that the latter result holds not only for discrete but also for continuous model weights and under a non-nested set-up. \cite{hansen_and_racine2012} show that the optimal weights can be obtained by minimizing the cross validation criterion, which allows for a more efficient use of data. Moreover, their approach allows to easily accommodate for heteroskedasticity. \cite{liu2015} points out that the asymptotic distribution of data dependent weights is non-standard, which complicates the inference. He augments the results from \cite{hjort_claeskens2003} and \cite{claeskens_hjort2008} and proposes a procedure that delivers asymptotically correct coverage probabilities for model averaging estimators. \cite{zhu_et_al2017} use a $J$-fold cross-validation criterion to construct optimal averaging weights for model averaging estimators under inequality constraints.

There is a large literature studying estimation under inequality constraints. \cite{andrews1999boundary} derives the asymptotic distribution of extremum estimators when the parameter of interest is on a boundary of the parameter space. His approach solves an asymptotically equivalent problem by minimizing a stochastic quadratic  objective function over a convex cone that approximates the parameter space. The approach follows \cite{chernoff1954}, \cite{feder1968}, \cite{pollard1985}, and \cite{wolak1989}. Andrews extends the results in these papers and allows for cases when the estimator objective function is undefined in the neighborhood of the true parameter.

There has been a growing interest in shrinkage estimators in the modern statistics literature. The main idea there is that shrinkage can be introduced through a penalty imposed on the estimator objective function. The most famous example is LASSO (\citealp{tibshirani1996}), which simultaneously shrinks and selects variables. Another seminal example is a Ridge regression, which shrinks the coefficients to zero, but does not perform selection. More complicated penalties lead to more interesting shrinkage spaces, e.g. a fused LASSO (\citealp{friedman_et_al2007}) can be used to shrink time-varying parameters towards random walk, i.e. it penalizes absolute time deviations of the form $|\theta_{t} - \theta_{t-1}|$. Another example is a nearly isotonic regression (\citealp{tibshirani2011nearly}) which shrinks the sequence of points towards a monotone sequence, i.e. it penalizes only positive part deviations $(\theta_{i} - \theta_{i+1})_{+}$. 

The remainder of the paper is organized as follows. Section \ref{sec:model} presents the general framework, describes the choice of shrinkage direction  and the local asymptotic framework. Section \ref{sec:estimation} introduces the inequality constrained shrinkage estimator. Section \ref{sec:distribution} derives the asymptotic distribution of the estimator. Section \ref{sec:asymptotic_risk} presents the risk dominance result. Section \ref{sec:weight} provides a feasible estimator for the data-dependent weight. Section \ref{sec:mc} demonstrates the finite sample performance of the ICSE in a series of simulations. In Section \ref{sec:slutsky} we apply the method to estimate gasoline demand under the Slutsky restriction. Section \ref{sec:conclusion} concludes. All the mathematical proofs and additional details are left to the Appendix.

\vspace{1em} 
We use the following notation throughout the paper: $\mathcal{I}_{n}$ denotes an $n \times n$ identity matrix. $\mathds{1}\{x \geq a\}$ is the indicator function that equals to one if $x \geq a$ and zero otherwise. We use $(x)_{+} = \max\{0,\,x\}$ to denote the ``positive part'' function. Finally, if $x$ is a vector, we use $x > a$ to denote each vector entry being strictly greater than $a$, the same holds for $x < a$.

\section{Model} \label{sec:model}

Suppose we observe a random array $\bm{X}_{n} = \{X_{in}\}_{i=1}^{n}$ of iid realizations. Let $Q_n(\theta)$ denote an extremum estimator objective function that depends on $\bm{X}_{n}$, for example, a GMM criterion or log likelihood function. The objective function is indexed by a parameter $\theta \in \Theta \subset \mathbb{R}^m$. 

The goal is to estimate the parameter of interest $\theta$ in a setting augmented by the belief that the true value of $\theta$ may be close (in a sense to be made clear later) to a restricted parameter space $\Theta_{0} \subset \Theta$ defined by a parametric restriction 
\begin{equation} \label{eq:restricted_set}
	\Theta_{0} = \{ \theta \in \Theta : r(\theta) \geq 0 \},
\end{equation}
where $r(\theta)$ is a differentiable function that maps $\mathbb{R}^{m} \rightarrow \mathbb{R}^{p}$. Let $R(\theta)$ denote the derivative $\frac{\partial}{\partial \theta'} r(\theta)$.

The pivotal point is that the true parameter value $\theta_{0}$ may not satisfy the restrictions, i.e. $\theta_{0}$ does not necessarily lie within $\Theta_{0}$. The restriction can be rather treated as a reasonable belief or ``prior'' about the likely value of $\theta_{0}$. It means that the empirical implication of the imposed theoretical restrictions are only ``approximately correct''.

\begin{remark}
	\textnormal{In this paper I focus on the parameter $\theta$ itself, the presented theory can be extended to functions of $\theta$ using the delta-method approach. However, one has to be cautious, since the level of shrinkage depends on the dimension of the function's output.}
\end{remark}

A common example is sign restrictions on all parameters, i.e. $p = m$. In this case the restricted space is $\Theta_{0} = \{ \theta \in \Theta : \theta \geq 0 \}$, where $r(\theta) = \theta$ and $R$ is simply an $m \times m$ identity matrix. The researcher may want to impose sign restrictions only on a subset of parameters. We can easily allow for that by partitioning the parameter space
\begin{equation*}
	\theta = 
	\begin{pmatrix}
		\theta_1 \\ \theta_2
	\end{pmatrix}
	\quad \quad
	\begin{matrix}
		m - p \\ p
	\end{matrix}
\end{equation*}  
then the sign restrictions take the form $r(\theta) = \theta_2$, and $R = [0_{p \times (m-p)} \,\vdots\, \mathcal{I}_{p}]$.

In general, $\Theta_{0}$ may be a non-linear subspace. This can be especially useful for structural estimation when an economic model implies non-linear inequality constraints on structural parameters.

\begin{example} \label{exmp:int_rate}
	\textnormal{In macroeconomics inequality restrictions often arise in estimation of DSGE models. \cite{moon2009} study an example of interest rate feedback rules, which we briefly describe here. Consider the following interest rate policy rule}
	\begin{equation} \label{eq:int_rate_rule}
		R_{t} = \rho_{R}R_{t-1} + (1 - \rho_{R})\psi_{1}\pi_{t} + (1 - \rho_{R})\psi_{2}x_{t} + \varepsilon_{R,\,t},
	\end{equation}
	\textnormal{where $R_{t}$ is the nominal interest rate in period t, $\pi_{t}$ is the inflation rate, and $x_{t}$ is a measure of real activity, such as output deviations from trend or output growth. The shock $\varepsilon_{R,\,t}$ captures unexpected deviations from the systematic component of the policy rule. To address potential endogeneity of both inflation and output in equilibrium, the researcher needs instrumental variables. Lagged variables of inflation and output are natural candidates. According to a large class of DSGE models, output does not fall in a response to an expansionary monetary shock, which leads to a moment restriction $\mathbb{E}[-x_{t}\varepsilon_{R,\,t}] \geq 0$.}
	
	\textnormal{One can estimate the model using the Generalized Method of Moments.\footnote{For the ease of exposition, we skip the details regarding the representation of a typical DSGE model and its solution.} Let $X_{t} = (R_{t-1},\,\pi_{t},\,x_{t})'$ be the vector of regressors, $Z_{t} = (R_{t-1},\,\pi_{t-1},\,x_{t-1})'$ be the vector of IVs, and $\theta = (\rho_{R},\,(1-\rho_{R})\psi_{1},\,(1-\rho_{R})\psi_{2})'$ be the parameter vector. Based on \eqref{eq:int_rate_rule}, one can form a finite sample moment condition $g_{t}(X_{t},\,Z_{t},\,R_{t};\theta) = T^{-1}\sum_{t=1}^{T} Z_{t}(R_{t} - X_{t}'\theta)$.}
	
	\textnormal{Instead of treating the moment restriction as an additional moment condition, one can impose it directly on the estimation problem. The finite sample analog is $-T^{-1}\sum_{t=1}^{T} x_{t}\varepsilon_{R,\,t} \geq 0$, or more explicitly,}
	\begin{equation*}
		\rho_{R}\sum_{t=1}^{T}x_{t}R_{t-1} + (1 - \rho_{R})\psi_{1}\sum_{t=1}^{T}x_{t}\pi_{t} + (1 - \rho_{R})\psi_{2}\sum_{t=1}^{T}x_{t}^{2} - \sum_{t=1}^{T}x_{t}R_{t} \geq 0,
	\end{equation*} 
	\textnormal{which imposes a linear inequality constraint on $\theta$.}
\end{example}

\begin{example} \label{exmp:slutzky}
	\textnormal{Inequality constraints also arise in many demand models. Consider a consumer who chooses her levels of consumption for different goods $j = 1,\dots,\,J$ by maximizing her utility function with respect to her budget constraint. One can show that the demand functions}
	\begin{equation*}
		D_{j} = D_{j}(p,\,m|\theta), \quad j = 1,\dots,\,J,
	\end{equation*}
	\textnormal{where $p$ is a price vector, $m$ is income, and $\theta$ are the structural parameters of interest, are not arbitrary. In particular, they must satisfy the budget constraint}
	\begin{equation*}
		\sum_{j=1}^{J}p_{j}D_{j}(p,\,m|\theta) = m.
	\end{equation*}
	\textnormal{Furthermore, since they solve a constrained optimization problem, they must satisfy the Slutsky matrix conditions. Let $S$ denote the Slutsky substitution matrix of size $J \times J$, whose generic entry is}
	\begin{equation*}
		S_{kj} = \frac{\partial D_{j}(p,\,m|\theta)}{\partial p_{k}} + \frac{\partial D_{j}(p,\,m|\theta)}{\partial m} D_{k}(p,\,m|\theta).
	\end{equation*}
	\textnormal{Economic theory tells us that such a matrix must be symmetric and negative semidefinite. These conditions imply inequality restrictions on the vector of structural parameters $\theta$.}
\end{example}

To measure the accuracy of an estimator $T_{n} = T_{n}(\bm{X}_{n})$ of $\theta$ we will use a known loss function $\ell(\theta,\,T_{n})$. The corresponding risk is just the expected loss
\begin{equation} \label{eq:risk_fun}
	R(\theta,\,T_{n}) = \mathbb{E}_{\theta} \ell(\theta,\,T_{n}).
\end{equation}
The most popular loss function in the literature is weighted quadratic loss, 
\begin{equation} \label{eq:quad_loss}
	\ell(\theta,\,T_{n}) = (T_{n} - \theta)'W(T_{n} - \theta) 
\end{equation}
for some weight matrix $W > 0$. The risk associated with \eqref{eq:quad_loss} is simply weighted mean squared error. In general, the choice of a loss function can be motived by an economic application, see \cite{hansen2016} for more examples.

The choice of a loss function plays a crucial role in the shrinkage estimator's behavior since the weights depend on the loss between the unrestricted and the restricted estimators. We specify the following regularity conditions for the loss function.

\begin{assumption} \label{loss_fun}
	The loss function $\ell(\theta,\,T_{n})$ satisfies
	\begin{itemize}
		\item [(a)] $\ell(\theta,\,T_{n}) \geq 0$
		\item [(b)] $\ell(\theta,\,\theta) = 0$
		\item [(c)]$W(\theta) = \left. \frac{1}{2}\frac{\partial^2}{\partial T_{n} \partial T_{n}'} \ell(\theta,\,T_{n}) \right\vert_{T_{n} = \theta}$ is continuous in a neighborhood of $\theta_{0}$.
	\end{itemize}
\end{assumption}

Assumptions \ref{loss_fun}(a) and (b) are standard properties of any loss function. The dominance result of \cite{james_stein1961} hinges on the quadratic loss function, however, our results hold for a more general family of loss functions. Assumption \ref{loss_fun}(c) requires the loss function $\ell(\theta,\,T_{n})$ to have a second derivative with respect to the second argument. This allows for smooth loss functions, like quadratic loss, and excludes non-smooth loss functions, such as absolute value loss.

The choice of a weight matrix also plays an important role. If one sets $W = \mathcal{I}_{m}$, \eqref{eq:quad_loss} becomes unweighted quadratic loss, which is appropriate for cases where all parameters are roughly identically scaled. However, when it is not the case, a weight matrix that renders a loss function which is robust to rotations of the parameter vector $\theta$ is a more plausible choice. We can fulfill the latter task by setting $W = \Omega^{-1}$, where $\Omega^{-1}$ is the inverse of the asymptotic variance of the unrestricted estimator. 

\subsection{Shrinkage direction} \label{sec:shr_dir}

The restriction in \eqref{eq:restricted_set} defining the direction of shrinkage, it is the main building block for the construction of our shrinkage estimator. Inequality constraints impose milder restrictions compared to equality constraints, which makes them harder to deal with. Equality restrictions provide the researcher with a particular shrinkage direction, however, with inequality constraints the shrinkage direction depends on the boundary which the true parameter value is close to. This stems from the properties of the inequality constrained estimator, see Section \ref{sec:distribution} for more details.

The researcher usually believes that restrictions are a reasonable simplification of the unrestricted model specification. And it is well known that if restrictions are correct, the restricted estimator renders more efficient estimates. In contrast, if not, the restricted estimates will be biased. In case of equality restrictions, the researcher can easily test them, however, testing inequality restrictions is an onerous task. Rather than testing inequality constraints, we can use them to construct an Inequality Constrained Shrinkage Estimator, and thereby, improve the efficiency of estimates. 

\subsection{Local asymptotic framework}

Our estimation framework is based on the belief that the empirical implications of theoretical restrictions are approximately correct. Put differently, it means that the parameter of interest $\theta_{0}$ does not necessarily lie within the restricted space $\Theta_{0}$, but is localized to it. We model that by assuming that the constraints are local to zero, i.e. $r(\theta_{0}) = cn^{-1/2}$, where $c \in \mathbb{R}^{p}$. In this framework $c$ is a slackness, or localizing, parameter which measures the discrepancy between $\theta_{0}$ and $\Theta_{0}$. When $c > 0$, then the constraints are satisfied and not binding, while if $c < 0$, the constraints are violated.\footnote{We are particularly interested in cases when constraints are locally violated. However, the analysis does not depend on the sign of the localizing parameter.} This modeling assumption ensures that the normalized asymptotic distribution of the ICSE is identical to its finite sample distribution under exact normality (see e.g. \citealp{hansen2016}).

We do not consider distant alternatives of the form $r(\theta_{0}) = \kappa_{n}c$, where $\kappa_{n}$ is $O(n^{-b})$ with $b < 1/2$, since we are interested in the asymptotic distribution of the normalized estimator. For simplicity, assume we have one only constraint. If $c < 0$, then $n^{1/2}r(\theta_{0}) = n^{1/2}\kappa_{n}c \rightarrow -\infty$, meaning that the constraint is violated, and we are better off with the restricted estimator. In contrast, if $c > 0$, then  $n^{1/2}r(\theta_{0}) = n^{1/2}\kappa_{n}c \rightarrow \infty$, meaning that the constraint is satisfied as a strict inequality, and we should resort to the unrestricted estimator.

\section{Estimation} \label{sec:estimation}

In order to define the shrinkage estimator, we first need to introduce unrestricted and restricted estimators.

The unrestricted estimator $\hat{\theta}_{n}$ of $\theta$ maximizes the objective function over $\theta \in \Theta$
\begin{equation*}
	Q_{n}(\hat{\theta}_{n}) = \sup_{\theta \in \Theta} Q_{n}(\theta).\footnote{We can allow for a numerical error by requiring $Q_{n}(\hat{\theta}_{n})$ to be within $o_{p}(1)$ of the global maximum of $Q_{n}(\theta)$, rather than the exact global maximum. This is a common assumption in the extremum estimators literature, yet fairly technical.}
\end{equation*} 
The restricted estimator $\tilde{\theta}_{n}$ is defined analogously 
\begin{equation*}
	Q_{n}(\tilde{\theta}_{n}) = \sup_{\theta \in \Theta_{0}} Q_{n}(\theta).
\end{equation*}
We assume that the maximum is unique so that $\hat{\theta}_{n}$ and $\tilde{\theta}_{n}$ are well-defined. 

The shrinkage estimator is defined as a weighted average of the unrestricted and restricted estimators
\begin{equation} \label{eq:shrink_est}
	\hat{\theta}^{*}_{n} = \hat{w}_{n} \hat{\theta}_{n} + (1 - \hat{w}_{n}) \tilde{\theta}_{n},
\end{equation}
where the weight is data driven and takes the form
\begin{equation} \label{eq:weight_def}
	\hat{w}_{n} = \left( 1 - \frac{\hat{\tau}_{n}}{n\ell(\hat{\theta}_{n},\,\tilde{\theta}_{n})} \right)_{+},
\end{equation}
where $\hat{\tau}_{n} \geq 0$ is the shrinkage parameter which controls the degree of shrinkage and $n\ell(\hat{\theta}_{n},\,\tilde{\theta}_{n})$ is the scaled loss between the unrestricted and restricted estimators. Under the quadratic loss, the latter becomes $n\left(\hat{\theta}_{n} - \tilde{\theta}_{n}\right)'W\left(\hat{\theta}_{n} - \tilde{\theta}_{n}\right)$.

The shrinkage parameter $\hat{\tau}_{n}$ is set to minimize the asymptotic risk of the ICSE. Thus, we allow $\hat{\tau}_{n}$ to be data-dependent and random, however, require it to converge in probability to a non-negative constant. 

\begin{assumption} \label{shrink_par}
	$\hat{\tau}_{n} \stackrel{p}{\rightarrow} \tau \geq 0$ as $n \rightarrow \infty$.
\end{assumption}

The degree of shrinkage determines an optimal bias variance tradeoff and depends on the ratio of the shrinkage parameter $\hat{\tau}_{n}$ to the loss $n\ell(\hat{\theta}_{n},\,\tilde{\theta}_{n})$. When the restricted estimator is very close to the unrestricted one, i.e. the loss is small, and $\hat{\tau}_{n} > n\ell(\hat{\theta}_{n},\,\tilde{\theta}_{n})$, we put all the weight on the restricted estimator, $\hat{w}_{n} = 0$ and $\hat{\theta}_{n}^{*} = \tilde{\theta}_{n}$. When $\hat{\tau}_{n} < n\ell(\hat{\theta}_{n},\,\tilde{\theta}_{n})$, then $\hat{\theta}_{n}^{*}$ is a weighted average of the restricted and unrestricted estimators. The larger the loss compared to the shrinkage parameter, the more weight we put on the unrestricted estimator. In other words, it means that if the regularization bias is small, we are better off trading it for a reduction in variance.

\section{Asymptotic distribution} \label{sec:distribution}

It is a well-known fact that the asymptotic distribution of the unrestricted extremum estimator is normal (see e.g. \cite{newey_mcfadden1994}), however, the asymptotic distribution of the inequality constrained estimator takes a more complicated form. Obtaining the restricted estimator requires solving an inequality constrained optimization problem, the solution to which depends on which constraints bind. As a result, the asymptotic distribution will take the form of a sum of truncated normal random variables.

We introduce the following regularity conditions.

\begin{assumption} \label{consistency}
	~\begin{itemize}
		\item [(a)] For some some function $Q(\theta): \Theta \rightarrow \mathbb{R}$, $\sup_{\theta \in \Theta} |Q_{n}(\theta) - Q(\theta)| \rightarrow_{p} 0$;
		\item [(b)] For all $\varepsilon > 0$, $\sup_{\theta \in \Theta / N(\theta_{0},\,\varepsilon)} Q(\theta) < Q(\theta_{0})$, where $N(\theta_{0},\,\varepsilon)$ is an $\varepsilon$-neighborhood of $\theta_{0}$.
	\end{itemize}
\end{assumption}

Assumption \ref{consistency}(a) ensures uniform convergence of the sample criterion function to the true criterion function. Assumption \ref{consistency}(b) requires the true criterion function to be uniquely maximized at $\theta_{0}$ in its neighborhood. These conditions guarantee that both the unrestricted and restricted estimators are consistent, i.e. $\hat{\theta}_{n} - \theta_{0}$ and $\tilde{\theta}_{n} - \theta_{0}$ are $o_{p}(1)$. Note that consistency does not depend on whether the estimator is restricted or not, the only thing that changes is the parameter space over which an estimator is defined (see e.g. Theorem 9.1 in \citealp{newey_mcfadden1994}).

\begin{assumption} \label{assumption_QA} 
	~\begin{itemize}
		\item [(a)] $\Theta$ is a compact subset of $\mathbb{R}^{m}$;
		\item [(b)] $\theta_{0}$ lies in the interior of $\Theta$;
		\item [(c)] $Q_{n}(\theta)$ is twice continuously differentiable in a neighborhood $N(\theta_0,\,\varepsilon)$ of $\theta$;
		\item [(d)] $n^{1/2} \frac{\partial}{\partial \theta} Q_{n}(\theta_{0}) \rightarrow_{d} G = \mathcal{N}(0,\,\mathcal{V})$ for some nonrandom positive definite matrix $\mathcal{V}$;
		\item [(e)] For $\theta \in N(\theta_{0},\,\varepsilon)$ there exists $\mathcal{J}(\theta)$ that is continuous and non-singular at $\theta_{0}$ and \\ $\sup_{\theta \in N(\theta_{0},\,\varepsilon)} \parallel \frac{\partial^{2}}{\partial \theta \partial \theta'} Q_{n}(\theta) - \mathcal{J}(\theta)\parallel \rightarrow_{p} 0$.
	\end{itemize}
\end{assumption}
Assumption \ref{assumption_QA} is a standard set of assumptions to ensure asymptotic normality of extremum estimators (see e.g. \citealp{newey_mcfadden1994}). Note that Assumption \ref{assumption_QA}(b) does not imply that $\theta_{0}$ lies in the interior of the restricted set $\Theta_{0}$, and whether $\theta_{0}$ belongs to the interior of $\Theta_{0}$ or not will affect the asymptotic distribution of both the restricted and shrinkage estimators. 

\begin{assumption} \label{constr_assumption}
	~\begin{itemize}
		\item [(a)] $R(\theta)$ is continuous in some neighborhood of $\theta_{0}$;
		\item [(b)] $R(\theta_{0})$ has full row rank.
	\end{itemize}
\end{assumption}
Assumption \ref{constr_assumption}(a) allows for applying the continuous mapping theorem, and Assumption \ref{constr_assumption}(b) rules out linearly dependent constraints.

\subsection{Solving an asymptotically equivalent problem} \label{subsec:asy_equiv_problem}

The asymptotic behavior of the unrestricted estimator is easily characterized, however, the distribution of the inequality constrained estimator is more complicated. Recall that in order to obtain the restricted estimator, we have to solve the following problem
\begin{equation} \label{eq:original_problem}
	\sup_{\theta \in \Theta} Q_{n}(\theta) \quad \text{s.t.} \quad r(\theta) \geq 0.
\end{equation}
 Dealing with non-linear inequality constrained optimization problems typically leads to very cumbersome calculations of the first order conditions. However, it turns out that we do not have to solve the original optimization problem. To derive the asymptotic distribution of the constrained estimator, it is sufficient to solve a simpler, asymptotically equivalent problem (see e.g. Section 21.3.2 in \citealp{gourieroux_monfort1995}). 

In our asymptotic analysis we follow \cite{andrews1999boundary} and rely on the quadratic approximation of the objective function around the true parameter value. In particular,
\begin{equation} \label{eq:quad_objective}
	Q_{n}(\theta) = Q_{nq}(\theta) + \xi_{n}(\theta).
\end{equation}
where
\begin{equation*}
	Q_{nq}(\theta) = Q_{n}(\theta_{0}) + \frac{\partial}{\partial \theta'} Q_{n}(\theta_{0}) (\theta - \theta_{0}) + \frac{1}{2}(\theta - \theta_{0})'\frac{\partial^{2}}{\partial \theta \partial \theta'} Q_{n}(\theta_{0}) (\theta - \theta_{0})
\end{equation*}
and $\xi_{n}(\theta)$ is the approximation error. We need to introduce some additional assumptions ensuring that $\xi_{n}(\theta)$ is of the right order, so that the estimator maximizing $Q_{nq}(\theta)$ has the same asymptotic distribution as of the true maximum. 
\begin{assumption} \label{stoch_diff}
	For all $\delta_{n} \rightarrow 0$, 
	\begin{equation*}
		\sup_{\theta \in \Theta: ||\theta - \theta_{0}|| \leq \delta_{n}} \frac{|\xi_{n}(\theta)|}{(1 + ||n^{1/2}(\theta - \theta_{0})||^{2})} = o_p(1).
	\end{equation*}
\end{assumption}
\cite{pollard1985} refers to Assumption \ref{stoch_diff} as stochastic differentiability, which is a weaker condition than $\xi_{n}(\theta)$ converging to 0 due to the presence of the denominator term $(1 + ||n^{1/2}(\theta - \theta_{0})||^{2})$. 

Let 
\begin{equation*}
	\mathcal{J}_{n} \equiv - \frac{\partial^{2}}{\partial \theta \partial \theta'} Q_{n}(\theta_{0}) \quad and \quad Z_{n} \equiv \mathcal{J}_{n}^{-1}n^{1/2} \frac{\partial}{\partial \theta} Q_{n}(\theta_{0}).
\end{equation*}
The quadratic approximation in \eqref{eq:quad_objective} can be rewritten as
\begin{align*}
	Q_{nq}(\theta) & = Q_{n}(\theta_{0}) + n^{-1/2} Z_{n}' \mathcal{J}_{n} (\theta - \theta_{0}) - \frac{1}{2} (\theta - \theta_{0})' \mathcal{J}_{n} (\theta - \theta_{0}) \\
	& = Q_{n}(\theta_{0}) + \frac{1}{2n} Z_{n}' \mathcal{J}_{n} Z_{n} - \frac{1}{n}q_{n}(n^{1/2}(\theta - \theta_{0})),
\end{align*}
where
\begin{equation*}
	q_{n}(\lambda) \equiv \frac{1}{2}(\lambda - Z_{n})' \mathcal{J}_{n} (\lambda - Z_{n}) \quad \text{and} \quad \lambda \in \mathbb{R}^{m}.
\end{equation*}
Note that under Assumption \ref{stoch_diff}, it is sufficient to minimize $q_{n}(n^{1/2}(\theta - \theta_{0}))$ to obtain a maximum of the quadratic approximation of $Q_{n}(\theta)$. When the parameter space is unrestricted, the estimator $\hat{\theta}_{n}$ equals to $\theta_{0} + n^{-1/2}Z_{n}$. Therefore, $n^{1/2}(\hat{\theta}_{n} - \theta_{0}) = Z_{n}$, and $Z_{n}$ determines the asymptotic distribution of the unrestricted estimator. A lemma below establishes the asymptotic distribution of the re-parameterized quadratic criterion function.  

\begin{lemma} \label{lemma:quad_approx}
	Under Assumptions \ref{consistency}--\ref{constr_assumption},
	
\begin{equation} \label{eq:crit_limit}
	\begin{aligned} 
		& \quad Z_{n} \hspace{0.1cm} \stackrel{d}{\rightarrow} \hspace{0.1cm} Z = \mathcal{J}^{-1}G, \\ 
		q_{n}(\lambda) \hspace{0.1cm} \stackrel{d}{\rightarrow} \hspace{0.1cm} &\; q(\lambda) \equiv \frac{1}{2}(\lambda - Z)'\mathcal{J}(\lambda - Z) \quad \forall \lambda \in \mathbb{R}^{m}.
	\end{aligned}
\end{equation}
\end{lemma}

Since the restricted estimator $\tilde{\theta}_{n}$ is consistent, its asymptotic distribution depends only on the features of the parameter space around the true parameter value $\theta_{0}$. We use the mean value expansion to approximate the constraints $r(\theta)$ around $\theta_{0}$,
\begin{equation*}
	r(\theta) = r(\theta_{0}) + R(\bar{\theta})(\theta - \theta_{0}) = c + R(\bar{\theta})n^{1/2}(\theta - \theta_{0})\geq 0,
\end{equation*}
where $\bar{\theta}$ lies on a segment between $\theta$ and $\theta_{0}$.\footnote{Essentially this approach is the same as approximating the restricted space by a cone of tangents (see e.g. \citealp{chernoff1954}, \citealp{feder1968}, and \citealp{andrews1999boundary}).} Since $\bar{\theta}_{n}$ lies on a segment between $\tilde{\theta}_{n}$ and $\theta_{0}$, under Assumptions \ref{consistency} and \ref{constr_assumption}, $R(\bar{\theta}_{n}) = R(\theta_{0}) + o_{p}(1)$. Let $R \equiv R(\theta_{0})$. As shown in Lemma \ref{lemma:distr_re} below, the asymptotic distribution of $n^{1/2}(\hat{\theta}_{n} - \theta_{0})$ is given by the distribution of
\begin{equation} \label{eq:reparam_problem}
	\begin{aligned}
		\tilde{\lambda} = \argmin_{\lambda \in \Lambda_{c}}\;  q(\lambda) 
	\end{aligned}
\end{equation}
where $\Lambda_{c} \equiv \{\lambda \in \mathbb{R}^{m}: c + R\lambda \geq 0\}$. By approximating the objective function with a quadratic counterpart and linearizing the constraints, we collapsed a potentially highly non-linear problem \eqref{eq:original_problem} to a simple quadratic programming problem.

There are $p$ inequality constraints which form $2^{p}$ different possible combinations of binding and non-binding constraints.\footnote{One can think of these combinations as possible boundaries of the restricted parameter space $\Theta_{0}$.} For each such combination the asymptotic distribution of the restricted estimator is simply a projection of the asymptotic limit of the unrestricted estimator $Z$ on the corresponding boundary. This is exactly the intuition in \cite{andrews1999boundary}, where he shows that under the standard asymptotics the asymptotic distribution of the extremum estimator, when the true parameter value is on a boundary, depends on binding constraints.

Let us introduce some notation simplifying the exposition. Let $L(\iota)$ be a linear subspace of the form $L(\iota) \equiv \{ l \in \mathbb{R}^{m}: c_{\iota} + R_{\iota}l = 0 \}$, where $\iota = 1,\dots,\,2^{p}$ represents one of the possible combinations of binding constraints. Let $\iota = 1$ denote the case when none of the constraints bind. Let $R_{\iota}$ consist of the rows of the Jacobian matrix $R$ corresponding to binding constraints indexed by $\iota$. By analogy, $\tilde{\mu}_{n,\iota}$ denotes a sub-vector of $\tilde{\mu}_{n}$  with entries corresponding to binding constraints indexed by $\iota$. Note that we also have to index the slackness parameter, as only the entries corresponding to binding constraints $c_{\iota}$ will affect the asymptotic distribution.

\begin{lemma} \label{lemma:distr_re}
	Suppose that Assumptions \ref{consistency}--\ref{stoch_diff} hold. Then, the asymptotic distribution of the constrained estimator takes the form
	\begin{equation} \label{eq:lambda_bind_gen}
		n^{-1/2}(\tilde{\theta}_{n} - \theta_{0}) \hspace{0.1cm} \stackrel{d}{\rightarrow} \hspace{0.1cm} \tilde{\lambda} \equiv Z - \sum_{\iota=2}^{2^{p}} P_{L(\iota)}(Z + h_{\iota}) \mathds{1}\{ \tilde{\mu}_{\iota} > 0, \; \tilde{\mu}_{-\iota} \leq 0 \},
	\end{equation}
	where $\tilde{\mu} = -(R\mathcal{J}^{-1}R')^{-1}(RZ + c)$ is the vector of Kuhn-Tucker multipliers for problem \eqref{eq:reparam_problem},
	\begin{equation}
		P_{L(\iota)} \equiv \mathcal{J}^{-1}R_{\iota}'\left(R_{\iota}\mathcal{J}^{-1}R_{\iota}'\right)^{-1}R_{\iota}
	\end{equation}
	is the projection on the linear subspace $L(\iota)$, $h_{\iota} \equiv R_{\iota}^{-1}c_{\iota}$ is the re-parameterized slackness parameter, and $R_{\iota}^{-1}$ is the right inverse of $R_{\iota}$.
\end{lemma}

Note that the distribution in \eqref{eq:lambda_bind_gen} is non-normal and depends on the re-parametarized slackness parameter $h$. The distribution takes the form of a sum of truncated normal random variables. Notice that the indicator functions are random: they depend on the asymptotic distribution of the Kuhn-Tucker multipliers.

 The slackness parameter enters the distribution through both the asymptotic bias term $P_{L(\iota)}h_{\iota}$ and the distribution of the Kuhn-Tucker multipliers $\tilde{\mu}$. From \eqref{eq:crit_limit} it follows that if $h_{j} \rightarrow \infty$, then $\tilde{\mu}_{j} \rightarrow -\infty$, implying that the $j^{th}$ constraint is not binding. If, on the contrary, $h_{j} \rightarrow -\infty$, then $\tilde{\mu}_{j} \rightarrow \infty$, resulting into the $j^{th}$ constraint being binding. 

The summation starts from $\iota = 2$ since we do not have to project the unrestricted estimator on any subspace when none of the constrains bind. Despite the seemingly complex expression, the basic intuition behind this formula is surprisingly simple. The asymptotic distribution of the inequality constrained estimator is just a projection of the asymptotic limit of the unconstrained estimator onto a boundary defined by the corresponding set of binding constraints. %

The following theorem summarizes the analysis above and presents the asymptotic distributions of the unrestricted, restricted, and shrinkage estimators.

\begin{theorem} \label{thm:asy_distr}
	Under Assumptions \ref{loss_fun}--\ref{stoch_diff}, 
	\begin{align}
		& n^{1/2}(\hat{\theta}_{n} - \theta_{0}) \hspace{0.1cm} \stackrel{d}{\rightarrow} \hspace{0.1cm} Z \sim \mathcal{N}(0,\,\Omega), \label{eq:ur_distr} \\
		& n^{1/2}(\tilde{\theta}_{n} - \theta_{0}) \hspace{0.1cm} \stackrel{d}{\rightarrow} \hspace{0.1cm} \tilde{\lambda} \equiv Z - \sum_{\iota=2}^{2^{p}} P_{L(\iota)}(Z + h_{\iota}) \mathds{1}\{ \tilde{\mu}_{\iota} > 0, \; \tilde{\mu}_{-\iota} \leq 0 \}, \label{eq:r_distr} \\
		& n\ell(\hat{\theta}_{n},\,\tilde{\theta}_{n}) \hspace{0.1cm} \stackrel{d}{\rightarrow} \hspace{0.1cm} \xi \equiv \sum_{\iota=2}^{2^{p}} (Z + h_{\iota})'P_{L(\iota)}' W P_{L(\iota)}(Z + h_{\iota}) \mathds{1}\{ \tilde{\mu}_{\iota} > 0, \; \tilde{\mu}_{-\iota} \leq 0 \}, \label{eq:loss_distr} \\
		& \hat{w}_{n} \hspace{0.1cm} \stackrel{d}{\rightarrow} \hspace{0.1cm} w = \left(1 - \frac{\tau}{\xi}\right)_{+}. \label{eq:weight_distr}
	\end{align}
	The asymptotic distribution of the inequality constrained shrinkage estimator is 
	\begin{align} \label{eq:shrink_distr}
		& n^{1/2}(\hat{\theta}_{n}^{*} - \theta_{0}) \hspace{0.1cm} \stackrel{d}{\rightarrow} \hspace{0.1cm} wZ + (1 - w)\tilde{\lambda}. \qquad\qquad\qquad\qquad\qquad\qquad\qquad
	\end{align}
\end{theorem}
\vspace{1 em}

\section{Asymptotic Risk} \label{sec:asymptotic_risk}

In practice obtaining the restricted estimator still requires solving a potentially complicated non-linear problem. This suggests that having an analytical closed form solution is extremely unlikely. Even if it is possible to derive an analytical solution, this solution will take a complex form, and the ICSE will inherit it. As a result, calculating its finite sample risk may  be infeasible. However, we know the asymptotic distribution of the ICSE, which means we can use the asymptotic risk to get a reasonable approximation of the finite sample risk.

Since the ICSE may not have a sufficient number of finite moments, to ensure existence we use an asymptotic trimmed loss. Let $T = \{T_{n}\}_{n=1}^{\infty}$ denote a sequence of estimators. The asymptotic risk of the estimator sequence $T$ is defined as
\begin{equation} \label{eq:asy_risk}
	\rho(h,\,T) = \lim_{\zeta \rightarrow \infty} \liminf_{n \rightarrow \infty} \mathbb{E}_{\theta_{0}} \min \left[ n\ell\left(\theta_{0},\,T_{n}\right),\,\zeta \right].
\end{equation}
The loss function is trimmed at $\zeta$, however, the trimming becomes negligible in large samples as $\zeta \rightarrow \infty$ with $n \rightarrow \infty$.

\cite{hansen2016} shows that whenever the loss function is locally quadratic, i.e. satisfies Assumption \ref{loss_fun}, the asymptotic risk, defined in \eqref{eq:asy_risk}, of an arbitrary estimator $T_{n}$, such that $n^{1/2}(T_{n} - \theta_{0}) \hspace{0.1cm} \stackrel{d}{\rightarrow} \hspace{0.1cm} \psi$, where $\psi$ is some random variable, can be calculated as
\begin{equation} \label{eq:quad_risk}
	\rho(h,\,T) = \mathbb{E}[\psi'W\psi].
\end{equation}

Equation \eqref{eq:quad_risk} allows us to calculate the asymptotic risk of the unrestricted and shrinkage estimators as expected weighted quadratic loss. Note that $n^{1/2}(\hat{\theta}_{n} - \theta_{0}) \hspace{0.1cm} \stackrel{d}{\rightarrow} \hspace{0.1cm} Z$, hence, the asymptotic risk of the unrestricted estimator is
\begin{equation} \label{eq:asy_risk_ur}
	\rho(h,\,\hat{\theta}_{n}) = \mathbb{E}[Z'WZ] = tr(W\mathbb{E}[ZZ']) = tr(W\Omega).
\end{equation}

Define an $m \times m$ matrix $A_{L(\iota)} \equiv W^{1/2\prime}\Omega P_{L(\iota)}'W^{1/2}$, let $\phi_{\max}(A_{L(\iota)})$ denote its largest eigenvalue. 

The following theorem establishes the main result of the paper.

\begin{theorem} \label{thm:risk_bound}
	Under Assumptions \ref{loss_fun}--\ref{stoch_diff}, if 
	\begin{equation} \label{eq:tau_bounds}
		0 < \tau \leq \sum_{\iota = 2}^{2^{p}} 2\left(tr(A_{L(\iota)}) - 2\phi_{\max}(A_{L(\iota)})\right) \gamma_{\iota},
	\end{equation}
	where
	\begin{equation} \label{eq:weights}
		\gamma_{\iota} \equiv \frac{\mathbb{E}[\xi_{L(\iota)}^{-1}]\mathbb{P}(\tilde{\mu}_{\iota} > 0,\,\tilde{\mu}_{-\iota} \leq 0)}{\sum_{\iota = 2}^{2^{p}}\mathbb{E}[\xi_{L(\iota)}^{-1}]\mathbb{P}(\tilde{\mu}_{\iota} > 0,\,\tilde{\mu}_{-\iota} \leq 0)},
	\end{equation}
	then for any $h$
	\begin{equation} \label{eq:risk_opt_bound}
		\rho(h,\,\hat{\theta}^{*}_{n}) < \rho(h,\,\hat{\theta}_{n}).
	\end{equation}

\end{theorem}
\vspace{1 em}

Equation \eqref{eq:risk_opt_bound} shows that the ICSE has strictly lower asymptotic risk than that of the unrestricted estimator for all values of the slackness parameter $h$, given that the shrinkage parameter $\tau$ satisfies the restriction \eqref{eq:tau_bounds}. 

The explicit risk bound for the ICSE is
\begin{equation} \label{eq:explicit_risk_bound}
	\rho(h,\,\hat{\theta}^{*}_{n}) < tr(W\Omega) - \tau\sum_{\iota = 2}^{2^{p}} \mathbb{E} \left[ \frac{
	2\left(tr(A_{L(\iota)}) - 2\phi_{\max}(A_{L(\iota)})\right) - \tau}{\xi_{L(\iota)}} \right] \mathbb{P}(\tilde{\mu}_{\iota} > 0,\,\tilde{\mu}_{-\iota} \leq 0)
\end{equation}
Since the bound in \eqref{eq:explicit_risk_bound} is quadratic in the shrinkage parameter $\tau$, there exists a unique optimal level of shrinkage $\tau^{*}$ that minimizes this bound,
\begin{equation} \label{eq:opt_shrink}
	\tau^{*} = \sum_{\iota = 2}^{2^{p}} \left(tr(A_{L(\iota)}) - 2\phi_{\max}(A_{L(\iota)})\right) \gamma_{\iota}.
\end{equation}

From \eqref{eq:weights} it follows that $\gamma_{\iota} \rightarrow 0$ when either the probability of $\iota^{th}$ event $\mathbb{P}(\tilde{\mu}_{\iota} > 0,\,\tilde{\mu}_{-\iota} \leq 0)$ is close to zero, or the expected inverse loss $\mathbb{E}[\xi^{-1}_{L(\iota)}]$ is approaching zero. The optimal shrinkage parameter puts more weight on events that are more likely to happen and on events where the restricted parameter is close to the unrestricted one. The behavior of $\gamma_{i}$ is ambiguous when $\mathbb{P}(\tilde{\mu}_{\iota} > 0,\,\tilde{\mu}_{-\iota} \leq 0)$ goes to zero and $\mathbb{E}[\xi^{-1}_{L(\iota)}]$ approaches infinity. 

When $W = \Omega^{-1}$, \eqref{eq:tau_bounds} simplifies to
\begin{equation} \label{eq:canonical_tau_bounds}
	0 < \tau \leq 2\left(\sum_{\iota = 2}^{2^{p}}p_{\iota}\gamma_{\iota} - 2\right),
\end{equation}
which leads to
\begin{equation} \label{eq:canonical_opt}
	\tau^{*} = \sum_{\iota = 2}^{2^{p}} p_{\iota}\gamma_{\iota} - 2.
\end{equation}
When $W = \Omega^{-1}$, $tr(A_{L(\iota)}) = tr(W\Omega P_{L(\iota)}') = tr(P_{L(\iota)}') = p_{\iota}$, and $\phi_{\max}(A_{L(\iota)}) = 1$, which gives condition \eqref{eq:canonical_tau_bounds}. This restriction on the shrinkage parameter has the same form as the classical James-Stein condition, $0 < \tau \leq 2(m-2)$, where $m$ is the dimension of the parameter of interest. As long as $m > 2$, the James-Stein estimator will dominate the unrestricted estimator in terms of asymptotic risk. In case of the ICSE, condition \eqref{eq:canonical_tau_bounds} requires $\sum_{\iota = 2}^{2^{p}} p_{\iota}\gamma_{\iota} > 2$. This means that the ``expected'' number of binding constraints must be greater than two for the ICSE to dominate. Constraints that are more likely to bind tell us which boundary of the restricted parameter space we are shrinking to, i.e. they determine the direction of shrinkage.

\section{Data-dependent weights} \label{sec:weight}

As it is pointed out by \cite{hjort_claeskens2003}, model averaging (and shrinkage) optimal weights cannot be consistently estimated in the local asymptotic framework since localizing parameters are $O(n^{1/2})$. And the ICSE is not an exception. Since the weights $\{\gamma_{\iota}\}_{\iota=2}^{2^p}$ depend on the localizing parameter, which is unknown, the optimal shrinkage parameter in \eqref{eq:opt_shrink} is infeasible. Furthermore, the localizing parameter $h$, which is a transformation of the original localizing parameter $c$, cannot be consistently estimated under the local asymptotic framework.

The weights $\{\gamma_{\iota}\}_{\iota=2}^{2^p}$ depend on the localizing parameter through the Kuhn-Tucker multipliers. The distribution of the Kuhn-Tucker multipliers is given by 
\begin{equation*}
	\tilde{\mu} = -(R\mathcal{J}^{-1}R')^{-1}R(Z + h) = -(R\mathcal{J}^{-1}R')^{-1}(RZ + c) \sim \mathcal{N}(\Psi(c),\,\Xi),
\end{equation*}
where $\Psi(c) = -(R\mathcal{J}^{-1}R')^{-1}c$ and $\Xi = (R\mathcal{J}^{-1}R')^{-1}R\Omega R'(R\mathcal{J}^{-1}R')^{-1}$. We observe that the mean of $\tilde{\mu}$ depends on the localizing parameter, thus, the distribution cannot be consistently estimated, as well as the corresponding probabilities. As a result, the optimal shrinkage parameter is infeasible.

A common approach in the literature is to obtain an asymptotically unbiased estimator of the localizing parameter $c$ (see e.g. \citealp{liu2015}). In our case $\hat{c}_{n} = n^{1/2}r(\hat{\theta}_{n})$ is an asymptotically unbiased estimator of $c$. To see this, approximate $\hat{c}_{n}$ around the true parameter value $\theta_{0}$ using the first-order Taylor expansion,
\begin{equation*}
	\begin{aligned}
		\hat{c}_{n} = n^{1/2}r(\hat{\theta}_{n}) & = n^{1/2}r(\theta_{0}) + n^{1/2}R(\theta_{0})(\hat{\theta}_{n} - \theta_{0}) + o_p(1) \\
		& \hspace{0.1cm} \stackrel{d}{\rightarrow} \hspace{0.1cm} c + RZ \sim \mathcal{N}(c,\,R\Omega R').
	\end{aligned}
\end{equation*}
Note that without the normalization a simple plug-in estimator $r(\hat{\theta_{n}})$ is just $O_{p}(1)$.

We propose to use a plug-in estimator of the optimal shrinkage parameter, $\hat{\tau}_{n}^{*} \equiv \tau^{*}(\hat{c}_{n})$. We can replace $R$, $\mathcal{J}$, and $\Omega$ with their consistent estimators $\hat{R}_{n} = R(\hat{\theta}_{n})$, $\hat{\mathcal{J}}_{n}$, and $\hat{\Omega}_{n} = \hat{\mathcal{J}}_{n}^{-1}\hat{\mathcal{V}}_{n}\hat{\mathcal{J}}_{n}^{-1}$. A consistent weighting matrix estimate, $\hat{W}_{n}$, can either be constructed from a specific context (e.g. an identity matrix, $\hat{W}_{n} = \mathcal{I}_{n}$) or as the second derivative of the loss function, i.e. $\hat{W}_{n} = W(\hat{\theta}_{n})$. 

We can then estimate $\tau^{*}$ by
\begin{equation} \label{eq:feasible_icse}
	\hat{\tau}_{n}^{*} = \sum_{\iota=2}^{2^p} \left(tr(\hat{A}_{n,L(\iota)}) - 2\phi_{max}(\hat{A}_{n,L(\iota)})\right) \hat{\gamma}_{n,\iota},
\end{equation}
where $\hat{A}_{n,L(\iota)} = \hat{W}_{n}^{1/2\prime}\hat{\Omega}_{n}\hat{R}_{n,\iota}'(\hat{R}_{n,\iota}\hat{\mathcal{J}}_{n}^{-1}\hat{R}_{n,\iota}')^{-1}\hat{R}_{n,\iota}\hat{\mathcal{J}}_{n}^{-1}\hat{W}_{n}^{1/2\prime}$ and the weights are constructed as
\begin{equation} \label{eq:feasible_gamma}
	\hat{\gamma}_{n,\iota} = \frac{\hat{\mathbb{E}}^{-1}[\xi_{L(\iota)}]\hat{\mathbb{P}}(\tilde{\mu}_{\iota} > 0,\,\tilde{\mu}_{-\iota} \leq 0)}{\sum_{\iota = 2}^{2^{p}}\hat{\mathbb{E}}^{-1}[\xi_{L(\iota)}]\hat{\mathbb{P}}(\tilde{\mu}_{\iota} > 0,\,\tilde{\mu}_{-\iota} \leq 0)}.
\end{equation}
In general, $\xi_{L(\iota)}$ follows a generalized $\chi^{2}$ distribution, which makes estimating its first inverse moment an extremely onerous task.\footnote{For more details on the calculation of inverse moments of the generalized $\chi^2$ distribution see e.g. \cite{jones1986}.} Instead, we proxy $\hat{\mathbb{E}}[\xi_{L(\iota)}^{-1}]$ with $\hat{\mathbb{E}}^{-1}[\xi_{L(\iota)}]$, which tends to work well in practice. We can consistently estimate the expected loss $\mathbb{E}[\xi_{L(\iota)}]$ by 
\begin{equation*}
	\hat{\mathbb{E}}[\xi_{L(\iota)}] = n(\hat{\theta}_{n} - \tilde{\theta}_{n,\iota})'\hat{W}_{n}(\hat{\theta}_{n} - \tilde{\theta}_{n,\iota}),
\end{equation*}
where $\tilde{\theta}_{n,\iota}$ is the equality constrained estimator given the constraints indexed by $\iota$. Probability estimates $\hat{\mathbb{P}}(\tilde{\mu}_{\iota} > 0,\,\tilde{\mu}_{-\iota} \leq 0)$ are based on the feasible distribution of the Kuhn-Tucker multipliers $\mathcal{N}(\hat{\Psi}_{n},\,\hat{\Xi}_{n})$, where $\hat{\Psi}_{n} = -(\hat{R}_{n}\hat{\mathcal{J}}_{n}^{-1}\hat{R}_{n}')^{-1}\hat{c}_{n}$ and $\hat{\Xi}_{n} = (\hat{R}_{n}\hat{\mathcal{J}}_{n}^{-1}\hat{R}_{n}')^{-1}\hat{R}_{n}\hat{\Omega}_{n}\hat{R}_{n}'(\hat{R}_{n}\hat{\mathcal{J}}_{n}^{-1}\hat{R}_{n}')^{-1}$.

Note, $\{\hat{\gamma}_{n,\iota}\}_{\iota=2}^{2^{p}}$ are not consistent estimates, since they do not converge in probability to their corresponding true values. Instead, they converge in distribution to random limits, which implies that the plug-in estimator of the shrinkage parameter $\hat{\tau}_{n}^{*} \stackrel{p}{\not\rightarrow}\tau^{*}$. Thus, the proposed feasible estimator \eqref{eq:feasible_icse} is not optimal in the sense that it uses the feasible data-driven weight that does not converge in probability to the optimal one. As a result, the dominance over the unrestricted estimator is not guaranteed. Despite that, in the following sections we show that the feasible estimator works well in practice.

\section{Monte Carlo Study} \label{sec:mc}

We demonstrate the finite sample performance of the ICSE in the following numerical simulation. Consider a following linear model. For $i = 1,\dots,\,n$,
\begin{equation*}
	y_{i} = x'_{1i}\theta_{1} + x'_{2i}\theta_{2} + \varepsilon_{i}.
\end{equation*}
The regressors $x_{1i}$ and $x_{2i}$ are $k_1 \times 1$ and $k_2 \times 1$, respectively. The vector of regressors, $x_{i}$, is distributed $\mathcal{N}(0,\,\Sigma)$, where $\Sigma_{jj} = 1$ and $\Sigma_{jk} = 0.5$ for $j \neq k$, and the error term, $\varepsilon_{i}$, is $\mathcal{N}(0,\,1)$. The goal is to estimate marginal effects under the belief that $\theta$ may be close to $\Theta_{0} = \{\theta \in \mathbb{R}^{k_1+k_2}:\theta_{1} \geq 0,\,\theta_{2} = 0\}$. For simplicity, in estimation we use a quadratic loss function. 

Let $\hat{\theta}_{n}$ denote the unrestricted OLS with $\hat{\Omega}_{n}$ being a consistent estimate of its asymptotic covariance matrix of $n^{1/2}(\hat{\theta}_{n} - \theta_{0})$. Let $\tilde{\theta}_{n}$ be the restricted OLS under $\theta_{1} \geq 0$ and $\theta_{2} = 0$.

We compare the performance of five different estimators of $\theta$. The first is $\hat{\theta}_{n}$, the unrestricted OLS estimator. The second is $\tilde{\theta}_{n}$, the restricted OLS estimator. The third estimator is the generalized James-Stein estimator of \cite{hansen2016} 
\begin{equation*}
	\hat{\theta}^{JS}_{n} = \hat{w}_{n}\hat{\theta}_{n}, \quad \hat{w}_{n} = \left(1 - \frac{k_1 + k_2 - 2}{n\hat{\theta}_{n}'\hat{\Omega}_{n}^{-1}\hat{\theta}_{n}}\right)_{+},
\end{equation*}
which shrinks both $\theta_1$ and $\theta_2$ to zero. 

The fourth estimator is the Empirical Bayes estimator $\hat{\theta}_{n}^{EB}$, which assumes the truncated normal prior $\theta|\nu \sim \mathcal{N}(0,\,1/\nu)\mathds{1}\{\theta \geq 0\}$, where $\nu$ is a hyper parameter that tells us how much weight to put on $\theta$ being equal to zero.\footnote{Further details can be found in Appendix \ref{app:eb}.} The higher the value of $\nu$, the more concentrated is the prior around zero, hence, the more mass is put on zero. The motivation for this estimator comes from the fact that the James-Stein estimator can be represented as an Empirical Bayes estimator (see e.g. \citealp{efron_morris1972a}). 

The last estimator is the feasible ICSE, which takes the same form but with the weight
\begin{equation*}
	\hat{w}^{*}_{n} = \left(1 - \frac{\sum_{\iota=1}^{2^{k_1}}p_{\iota}\hat{\gamma}_{n,\iota} - 2}{n(\hat{\theta}_{n} - \tilde{\theta}_{n})'\hat{\Omega}_{n}^{-1}(\hat{\theta}_{n} - \tilde{\theta}_{n})}\right)_{+},
\end{equation*}
where $p_{\iota}$ is the total number of binding constraints in $\iota$ case. Note, since there are two equality constraints, if none of the inequality constraints bind, $\iota = 1$, $p_1 = 2$. The weights $\{\hat{\gamma}_{n,\iota}\}_{\iota=1}^{2^{k_1}}$ are estimated by \eqref{eq:feasible_gamma}.

The estimators are compared by the mean square error (MSE), which is calculated based on $N = 2,000$ replications. For the ease of exposition, we normalize the MSE of the unrestricted estimator to be equal to one so that the MSE of other estimators are given relative to the MSE of the unrestricted one.

We set the regression coefficients as $\theta_1 = (1,\,1,\,1,\,b,\dots,\,b)$, and $\theta_{2} = (c,\,c,\dots,\,c)'$. Thus, the remaining control parameters in the model are $k_1$, $b$, $c$, and $n$. The value of $b$ allows us to control the strength of the inequality constraints, i.e. whether they are satisfied or not, and $c$ controls the strength of the equality constraints. 

Note that the inequality constraints do not change simultaneously with $b$. When $b$ is negative, the first three constraints are satisfied, while the remaining $k_1 - 3$ constraints are violated. As a result, shrinking towards inequality constraints is fundamentally different from shrinking towards equality constraints.

\begin{figure}[!ht] 
	\centering
	\includegraphics[scale=0.685]{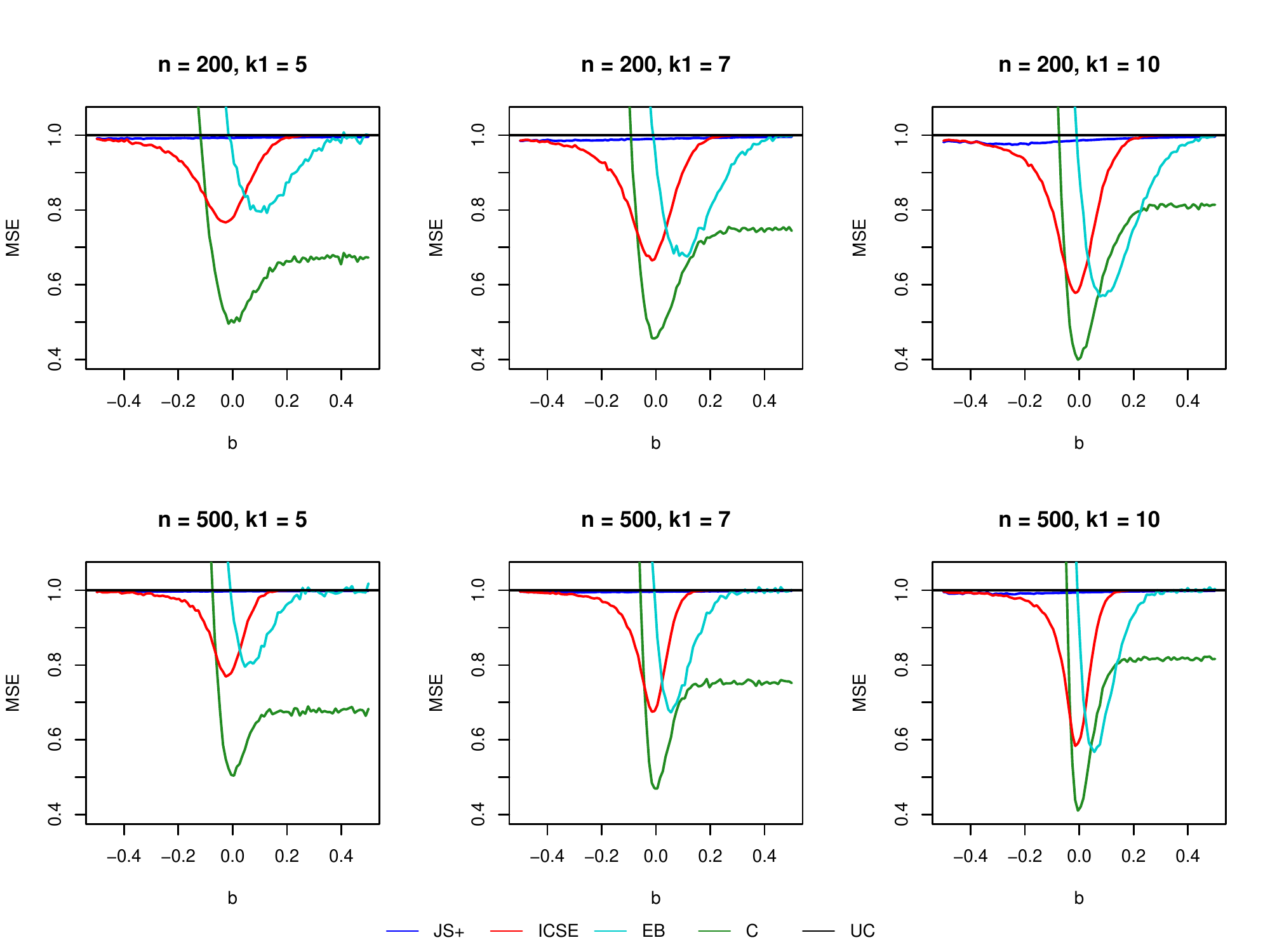}
	\mycaption{MC results}{This figure shows normalized MSEs for different combinations of $k_1 = \{5,\,7,\,10\}$ and $n = \{200,\,500\}$.}
	\label{fig:mse_lin2_no_ma}
\end{figure}

In Figure \ref{fig:mse_lin2_no_ma}, we display the results for  $n = \{200,\,500\}$, $k_1 = \{5,\,7,\,10\}$, and vary $b$ on a $100$-point equispaced grid from $-0.5$ to $0.5$. We set $c=0$ so that the equality constraints are satisfied.

First, the feasible ICSE dominates the unrestricted estimator, while the restricted estimator along with the EB estimator do worse than the unrestricted one when the constraints are violated. Since the posterior is truncated at zero, the EB estimates of $\theta_1$ are always positive, which explains the result. 

Second, we observe that the James-Stein estimator exhibits almost no improvement upon the unrestricted estimator. This behavior is expected since the James-Stein estimator shrinks all the constraints towards zero, which is fundamentally different from shrinking towards inequalities. As a result, the James-Stein estimator puts almost no weight on the restricted estimator. If the shrinkage direction is chosen poorly, it will lead to a large bias resulting into poor overall performance. Thus, the shrinkage gains are guaranteed only if the shrinkage direction is chosen properly. 

When $b < 0$, the restricted and EB estimators perform worse than the unrestricted one, while the feasible ICSE achieves significant MSE reduction gains. When $b$ approaches zero, the constrained estimator starts to dominate the feasible ICSE. When $b > 0$, the constrained estimator dominates both shrinkage estimators, however, the EB estimator achieves lower MSE when $b$ is slightly greater than zero. As $b$ grows, the EB estimator converges to the unrestricted estimator. When the number of observations increases, the prior gets less weight pushing the drop in the MSE closer to $b = 0$. Notice that the MSE of the restricted estimator does not converge to the one of the unrestricted. Since $c = 0$, the inequality constrained estimator is more accurate than the unrestricted one, which explains the result. Moreover, as the number of inequality constraints grows, the difference between the unconstrained and constrained estimators vanishes, resulting into lower MSE gains of the constrained estimator over the unconstrained one.

Finally, when the number of inequality constraints increases, the shrinkage effect of the feasible ICSE and EB estimators becomes more prominent, which supports the theoretical findings.

\section{Empirical Application: Demand Estimation under the Slutsky Restriction} \label{sec:slutsky}

In our empirical application we consider consumer demand estimation under the Slutsky restriction (see Example \ref{exmp:slutzky} for more details). In this application we build on literature on demand estimation under shape restrictions, especially on the recent results by \cite{blundell2012}, \cite{dette2016}, and \cite{blundell2017}.

Our goal is to estimate price and income elasticities of gasoline demand for different income levels. Slutsky condition is an inequality constraint on the demand function ensuring that the compensated own-price elasticities are negative. Despite the fact that in theory consumer choices should abide the Slutsky restriction, in the data we might find evidence suggesting otherwise. For example, if gasoline prices are too high and households  anticipate them to rise further, then households will tend to buy more gasoline now and store it for future use  resulting in positive compensated price elasticity, which violates the Slutsky restriction. That is exactly where we expect shrinkage gains. Implementation details can be found in Appendix \ref{app:ea_details}.

We use the same data and sample construction as \cite{blundell2017}, which we briefly describe here.\footnote{Further details on sample construction can be found in Section IV.A of \cite{blundell2017}. A more detailed description of the NHTS dataset is presented in Section 3 of \cite{blundell2012}.} The data are from the 2001 National Household Travel Survey (NHTS). The sample is constructed to reduce heterogeneity by restricting the analysis to households with a white respondent, two or more adults, at least one child under age 16, and at least one driver. Households in the most rural areas and in Hawaii are excluded from the sample, as well as are households with missing relevant variables or without a gasoline based vehicle. The resulting sample contains 3,640 observations, where the key variables of interest are gasoline demand, price of gasoline, and household income.

\begin{figure}[!ht]
	\centering
	\begin{subfigure}{0.75\textwidth}
		\includegraphics[width=\textwidth,height=6cm]{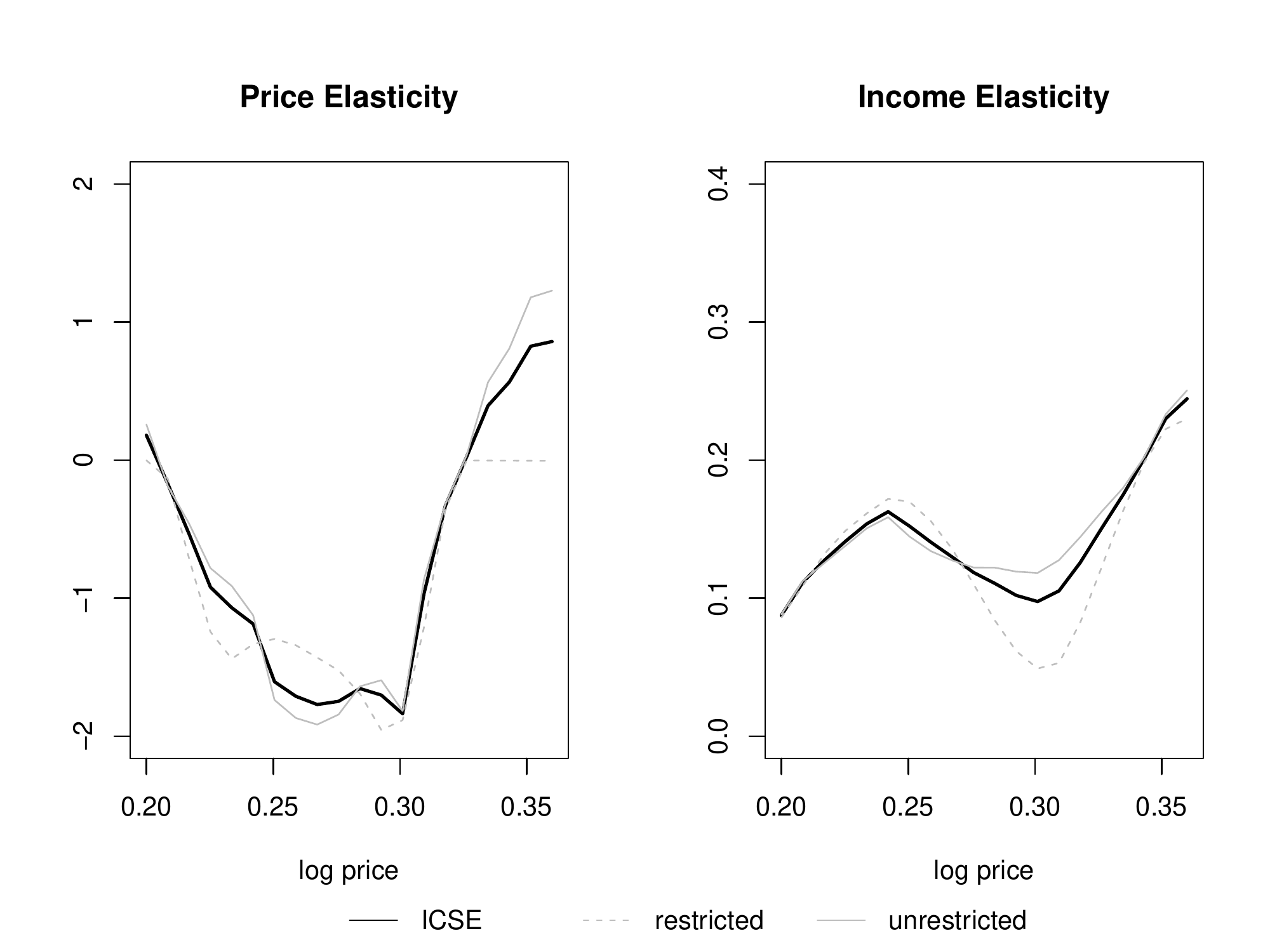}
		\caption{High Income}
		\label{subfig:icse_gas_high}
	\end{subfigure}
	
	\begin{subfigure}{0.75\textwidth}
		\includegraphics[width=\textwidth,height=6cm]{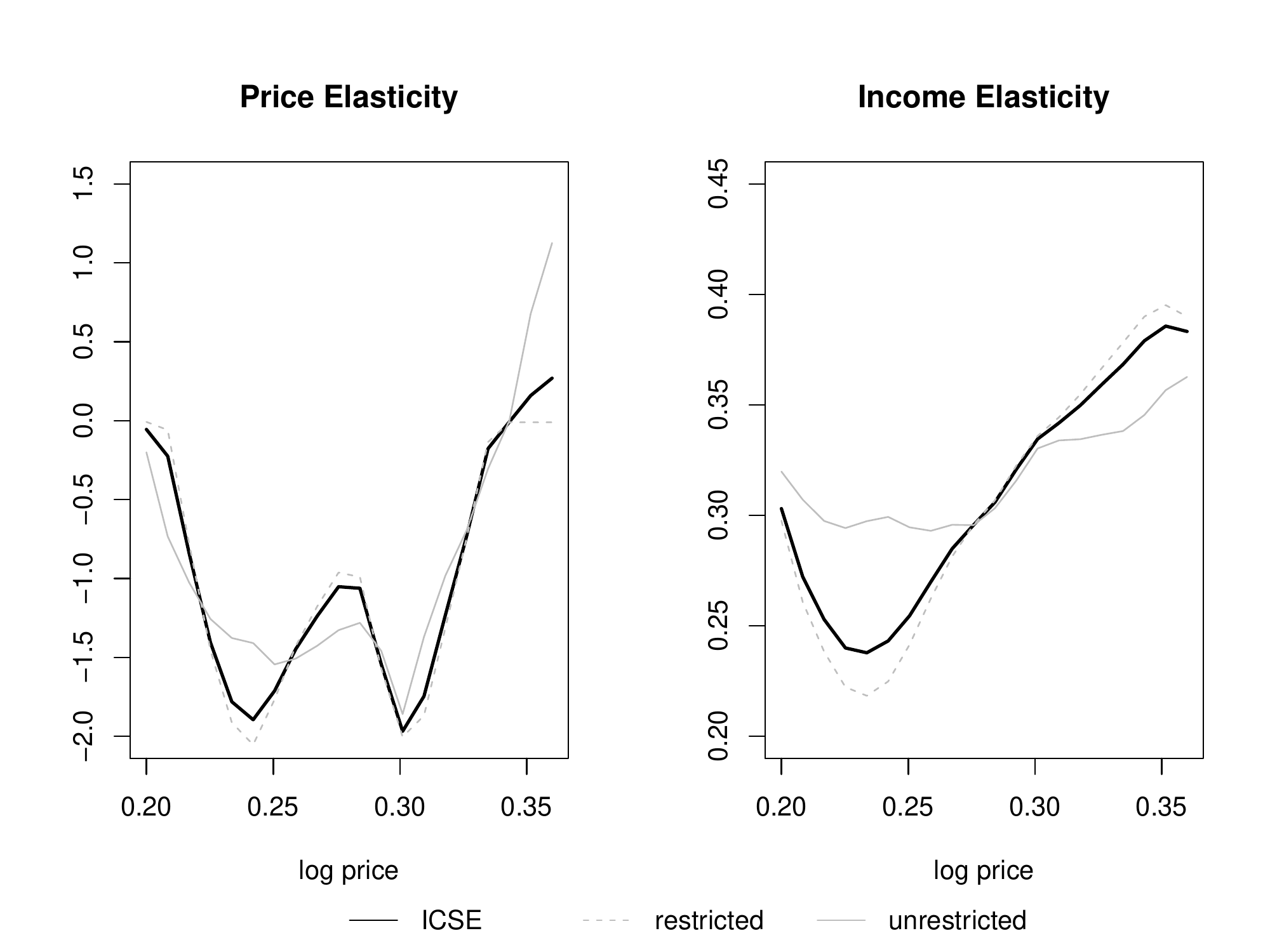}
		\caption{Medium Income}
		\label{subfig:icse_gas_med}
	\end{subfigure}
	
	\begin{subfigure}{0.75\textwidth}
		\includegraphics[width=\textwidth,height=6cm]{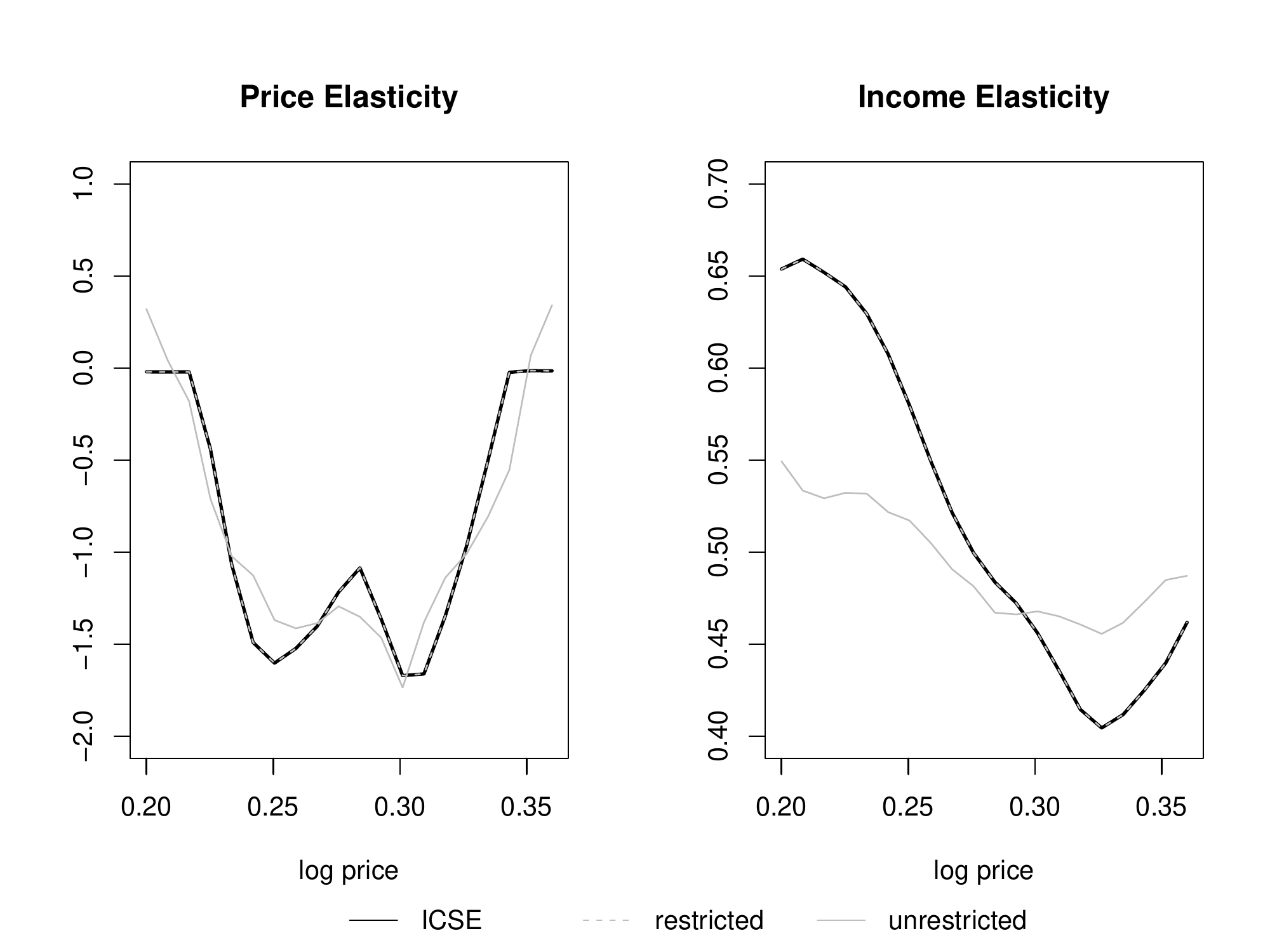}
		\caption{Low Income}
		\label{subfig:icse_gas_low}
	\end{subfigure}
	\vspace{1em}
	\mycaption{Price and income elasticity estimates}{This figure shows the unrestricted, restricted, and ICSE estimates of price and income elasticities.}
	\label{fig:icse_gas}
\end{figure}

We demonstrate estimates for low, medium, and high income level groups which correspond to the first, second, and third quartile, respectively. As a base estimator we use the local linear regression (LLR) with 20 grid points in the observe range of values for the log price. We set the bandwidth for log price and log income using the rule of thumb to their respective standard deviations. Further implementation details are left for Appendix \ref{app:ea_details}.

Figure \ref{fig:icse_gas} plots the unrestricted, restricted, and ICSE estimates of price and income elasticities as functions of price, across the income levels. Degree of shrinkage differs across income groups. We estimate the weight on the unrestricted estimator $\hat{w}$ to be 0 for low income group, $0.25$ for medium income group, and $0.75$ for high income group. Thus, consumers from higher income groups are more likely to have upward sloping demand curves, which is consistent with the results in \cite{blundell2012}. However, the Empirical Bayes estimates of \cite{kasy2018}, based on the local linear quantile regression, suggest to shrink more towards the restricted estimates for all income groups. The reason the estimates differ is due to the fact that ICSE shrinks all components of $\hat{\beta}$ by the same factor $\hat{w}$, while the EB estimator provides component-wise shrinkage with different shrinkage factors (for more details see Section 4.1 in \citealp{kasy2018}). 

\section{Conclusion} \label{sec:conclusion}

In this paper we have shown how to shrink extremum estimators towards theoretical restrictions in form of inequality constraints. The ICSE asymptotically uniformly dominates the unrestricted estimator. The shrinkage direction depends only on the binding constraints rendering it \textit{ex ante} unknown to the researcher, which is the main difference compared to shrinking towards equality constraints. 

An important caveat, however, is that due to the presence of localizing parameters that cannot be consistently estimated we cannot guarantee the risk dominance result in finite samples, which is a common problem in frequentist model averaging and shrinkage literatures. One possible improvement would be to establish uniform dominance of the ICSE, but we leave this for future research. 

\printbibliography

\newpage

\begin{appendices} 

\section{Lemmas and Proofs} \label{app:proofs}

\renewcommand{\theequation}{A.\arabic{equation}}
\setcounter{equation}{0}

\subsection{Proof of Lemma \ref{lemma:quad_approx}}

Assumptions \ref{consistency}--\ref{assumption_QA} along with the Slutsky lemma and continuous mapping theorem immediately give us $Z_{n} \rightarrow_{d} Z = \mathcal{J}^{-1}G$ and $q_{n}(\lambda) \rightarrow_{d} \frac{1}{2} (\lambda - Z)'\mathcal{J}(\lambda - Z)$. \qed \\

\subsection{Proof of Lemma \ref{lemma:distr_re}}

The proof follows directly from Theorems 3 and 5 in \cite{andrews1999boundary} with two slight modifications. First,  note that the restricted space $\Lambda_{c}$ in \eqref{eq:reparam_problem} is a convex cone with a (possibly) non-zero vertex $-c$, while in \cite{andrews1999boundary} the cone has a zero vertex. This changes the form of projection in \eqref{eq:lambda_bind_gen}, which in our case accommodates for the non-zero vertex. Second, the indicator functions in \eqref{eq:lambda_bind_gen} are given in terms of the Kuhn-Tucker multipliers (see the derivation below) instead of the asymptotic limits of the subvectors of $\tilde{\theta}_{n}$. \qed \\

Deriving Kunh-Tucker multipliers requires writing down the first order conditions\footnote{FOCs are both necessary and sufficient since it is a quadratic programming problem with linear constraints.} to \eqref{eq:reparam_problem}:
\begin{equation*} \label{eq:focs_restricted}
	\begin{aligned}
		&\mathcal{J}(\tilde{\lambda} - Z) - R'\tilde{\mu} = 0 \\
		&\tilde{\mu}'(c + R)\tilde{\lambda}) = 0 \\
		&\tilde{\mu} \geq 0, \quad c + R\tilde{\lambda} \geq 0,
	\end{aligned}
\end{equation*}
where the vector of Kuhn-Tucker multipliers satisfying the first order conditions is given by
\begin{equation*}
	\tilde{\mu} = -(R\mathcal{J}^{-1}R')^{-1}\left(RZ + c\right).
\end{equation*}

\subsection{Proof of Theorem \ref{thm:asy_distr}}

To prove \eqref{eq:loss_distr}, we begin by taking a second order mean value expansion of the loss function around $\hat{\theta}_{n}$,
\begin{equation*}
	n\ell(\hat{\theta}_{n},\,\tilde{\theta}_{n}) = 	n\ell(\hat{\theta}_{n},\,\hat{\theta}_{n}) + n\left.\frac{\partial}{\partial \theta'} \ell(\hat{\theta}_{n},\,\theta)\right\vert_{\theta = \hat{\theta}_{n}}(\tilde{\theta}_{n} - \hat{\theta}_{n}) + n(\tilde{\theta}_{n} - \hat{\theta}_{n})'W(\theta^{*}_{n})(\tilde{\theta}_{n} - \hat{\theta}_{n}),
\end{equation*}
where $\theta^{*}_{n}$ lies on a line segment between $\hat{\theta}_{n}$ and $\tilde{\theta}_{n}$. Assumption \ref{loss_fun}(b) implies that $n\ell(\hat{\theta}_{n},\,\hat{\theta}_{n}) = 0$. By Assumption \ref{loss_fun}(c) and the fact that $\ell(\hat{\theta}_{n},\,\hat{\theta}_{n})$ is minimized at $\hat{\theta}_{n}$, $\left.\frac{\partial}{\partial \theta'} \ell(\hat{\theta}_{n},\,\theta)\right\vert_{\theta = \hat{\theta}_{n}} = 0$. Consistency of both $\hat{\theta}_{n}$ and $\tilde{\theta}_{n}$ along with Assumption \ref{loss_fun}(c) implies that $W(\theta^{*}_{n}) = W + o_{p}(1)$. 
	
We have shown that the unrestricted estimator is asymptotically equal to $Z$. Combining this fact with
 the results from Lemma \ref{lemma:distr_re} gives
\begin{equation*}
	n^{-1/2}(\hat{\theta}_{n} - \tilde{\theta}_{n}) \rightarrow_{d} \sum_{\iota=2}^{2^{p}} P_{L(\iota)}(Z + h_{\iota}) \mathds{1}\{ \tilde{\mu}_{\iota} > 0, \; \tilde{\mu}_{-\iota} \leq 0 \}.
\end{equation*}
Hence, the asymptotic distribution of the loss function is 
\begin{align*}
	n\ell(\hat{\theta}_{n},\,\tilde{\theta}_{n}) = & n(\tilde{\theta}_{n} - \hat{\theta}_{n})'W(\theta^{*}_{n}(\tilde{\theta}_{n} - \hat{\theta}_{n}) \\
	& \rightarrow_{d} \sum_{\iota=2}^{2^{p}} (Z + h_{\iota})'P_{L(\iota)}' W P_{L(\iota)}(Z + h_{\iota}) \mathds{1}\{ \tilde{\mu}_{\iota} > 0, \; \tilde{\mu}_{-\iota} \leq 0 \} = \xi,
\end{align*}
which is \eqref{eq:loss_distr}.
	
\eqref{eq:weight_distr} and \eqref{eq:shrink_distr} follow by the continuous mapping theorem and Assumption \ref{shrink_par}. \qed \\

To derive the bound, we use a version of Stein's Lemma \parencite{stein1981} presented in \cite{hansen2016}.

\begin{lemma} \label{steins_lemma}
	If $Z \sim \mathcal{N}(0,\,\mathcal{V})$ is $m \times 1$, $K$ is $m \times m$, and $\eta(x):\mathbb{R}^{m} \rightarrow \mathbb{R}^{m}$ is absolutely continuous, then
	\begin{equation*}
		\mathbb{E}\left[\eta(Z + h)'KZ\right] = \mathbb{E}tr\left(\frac{\partial}{\partial x'}\eta(Z+h)K\mathcal{V}\right).
	\end{equation*} 
\end{lemma} 

\subsection{Proof of Theorem \ref{thm:risk_bound}}

The proof is similar to \cite{hansen2016}.
First, observe that $n^{1/2}(\hat{\theta}^{*}_{n} - \theta_{0}) \rightarrow \psi^{*}$, where $\psi^{*} = wZ + (1 - w)\tilde{\lambda}$, as shown in \eqref{eq:shrink_distr}. Hence, the risk of the shrinkage estimator can be calculated as $\rho(h,\,\hat{\theta}^{*}_{n}) = \mathbb{E}[\psi^{*\prime}W\psi^{*}]$. The distribution of the variable $\psi^{*}$ is based on the classic James-Stein distribution with positive part trimming. Define a similar random variable $\psi$ without positive part trimming
\begin{equation} \label{eq:js_no+}
	\psi = Z\left(1 - \frac{\tau}{\xi}\right) + \frac{\tau}{\xi}\tilde{\lambda} = Z - \frac{\tau}{\xi}\sum_{\iota = 2}^{2^{p}} P_{L(\iota)}(Z + h_{\iota}) \mathds{1}\{ \tilde{\mu}_{\iota} > 0, \; \tilde{\mu}_{-\iota} \leq 0 \}.
\end{equation}

It is a well-known fact that positive part trimming always reduces risk under the standard quadratic loss (see e.g. Theorem 5.5.4 in \cite{lehmann_casella1998}, or Lemma 2 in \cite{hansen2015}). Thus, using this fact and \eqref{eq:quad_risk},
\begin{equation} \label{eq:asy_risk_js}
	\rho(h,\,\hat{\theta}^{*}_{n}) = \mathbb{E}[\psi^{*\prime}W\psi^{*}] < \mathbb{E}[\psi'W\psi].
\end{equation}
Using \eqref{eq:js_no+}, we calculate that the asymptotic risk in \eqref{eq:asy_risk_js} is equal to
\begin{equation} \label{eq:risk_bound0}
	\begin{aligned}
		\mathbb{E}[\psi'W\psi] = & \mathbb{E}[Z'WZ] \\
		& + \tau^{2}\mathbb{E}\left[ \frac{(\sum_{\iota = 2}^{2^{p}} P_{L(\iota)}(Z + h_{\iota}) \mathds{1}\{ \tilde{\mu}_{\iota} > 0, \; \tilde{\mu}_{-\iota} \leq 0 \})'W(\sum_{\iota = 2}^{2^{p}} P_{L(\iota)}(Z + h_{\iota}) \mathds{1}\{ \tilde{\mu}_{\iota} > 0, \; \tilde{\mu}_{-\iota} \leq 0 \})}{\xi^{2}} \right] \\
		& -2\tau \mathbb{E}\left[ \frac{\sum_{\iota = 2}^{2^{p}} (Z + h_{\iota})'P_{L(\iota)}'WZ \mathds{1}\{ \tilde{\mu}_{\iota} > 0, \; \tilde{\mu}_{-\iota} \leq 0 \}}{\xi} \right] \\
		= & tr(W\Omega) + \tau^{2} \sum_{\iota = 2}^{2^{p}} \mathbb{E} \left[ \frac{1}{\xi_{L(\iota)}} \right] \mathbb{P}(\tilde{\mu}_{\iota} > 0, \; \tilde{\mu}_{-\iota} \leq 0) \\
		& - 2\tau \sum_{\iota = 2}^{2^{p}} \mathbb{E} \left[ \frac{(Z + h_{\iota})'P_{L(\iota)}'WZ}{\xi_{L(\iota)}} \right] \mathbb{P}(\tilde{\mu}_{\iota} > 0, \; \tilde{\mu}_{-\iota} \leq 0) \\
		= & tr(W\Omega) + \sum_{\iota = 2}^{2^{p}} \left( \tau^{2} \mathbb{E} \left[ \frac{1}{\xi_{L(\iota)}} \right] - 2\tau \mathbb{E} \left[ \frac{(Z + h_{\iota})'P_{L(\iota)}'WZ}{\xi_{L(\iota)}} \right] \right) \mathbb{P}(\tilde{\mu}_{\iota} > 0, \; \tilde{\mu}_{-\iota} \leq 0).
	\end{aligned}
\end{equation}
Have a closer look at the second expectation term of the $\iota^{th}$ summand,
\begin{equation*}
	\mathbb{E} \left[ \frac{(Z + h_{\iota})'P_{L(\iota)}'WZ}{\xi_{L(\iota)}} \right] = \mathbb{E} \left[ \eta_{L(\iota)}(Z + h_{\iota})'P_{L(\iota)}'WZ \right],
\end{equation*}
where
\begin{equation*}
	\eta_{L(\iota)}(x) = \frac{x}{x'B_{L(\iota)}x}.
\end{equation*}

Next, before applying the Stein's Lemma we calculate
\begin{equation} \label{eq:eta_deriv}
	\frac{\partial}{\partial x'} \eta_{L(\iota)}(x) = \left(\frac{1}{x'B_{L(\iota)}x}\right)\mathcal{I} - \frac{2B_{L(\iota)}xx'}{(x'B_{L(\iota)}x)^{2}}.
\end{equation}
Using Lemma \ref{steins_lemma} and \eqref{eq:eta_deriv},
\begin{equation} \label{eq:risk_bound1}
	\begin{aligned}
		 \mathbb{E} \left[ \eta_{L(\iota)}(Z + h_{\iota})'P_{L(\iota)}'WZ \right] & = \mathbb{E}tr\left( \frac{\partial}{\partial x'} \eta_{L(\iota)}(Z + h_{\iota})P_{L(\iota)}'W\Omega \right) \\
		 & = \mathbb{E}tr\left( \frac{P_{L(\iota)}'W\Omega}{(Z + h_{\iota})'B_{L(\iota)}(Z + h_{\iota})} \right) \\
		 & - 2\mathbb{E}tr\left( \frac{B_{L(\iota)}(Z + h_{\iota})(Z + h_{\iota})'P_{L(\iota)}'W\Omega}{[(Z + h_{\iota})'B_{L(\iota)}(Z + h_{\iota})]^{2}} \right).
	\end{aligned}
\end{equation}
Moreover, 
\begin{equation} \label{eq:risk_bound2}
	\begin{aligned}
		tr\left(B_{L(\iota)}(Z + h_{\iota})(Z + h_{\iota})'P_{L(\iota)}'W\Omega\right) & = (Z + h_{\iota})'P_{L(\iota)}'W\Omega B_{L(\iota)}(Z + h_{\iota}) \\
		& = (Z + h_{\iota})'P_{L(\iota)}'W^{1/2}W^{1/2\prime}\Omega P_{L(\iota)}'W^{1/2}W^{1/2\prime}P_{L(\iota)}(Z + h_{\iota}) \\
		& = (Z + h_{\iota})'\tilde{B}_{L(\iota)}'A_{L(\iota)}\tilde{B}_{L(\iota)}(Z + h_{\iota})
	\end{aligned}
\end{equation}
where
\begin{equation*}
	\tilde{B}_{L(\iota)} = W^{1/2\prime}P_{L(\iota)}, \quad \text{and} \quad B_{L(\iota)} = \tilde{B}_{L(\iota)}'\tilde{B}_{L(\iota)}.
\end{equation*}
Combining \eqref{eq:risk_bound1}, \eqref{eq:risk_bound2}, and the fact that $tr(A_{L(\iota)}) = tr(P_{L(\iota)}'W\Omega)$, we get
\begin{equation} \label{eq:risk_bound3}
	\begin{aligned}
		 \mathbb{E} \left[ \eta_{L(\iota)}(Z + h_{\iota})'P_{L(\iota)}'WZ \right] 
		 & = \mathbb{E}\left( \frac{tr(A_{L(\iota)})}{(Z + h_{\iota})'B_{L(\iota)}(Z + h_{\iota})} \right) - 2\mathbb{E}\left( \frac{(Z + h_{\iota})'\tilde{B}_{L(\iota)}'A_{L(\iota)}\tilde{B}_{L(\iota)}(Z + h_{\iota})}{[(Z + h_{\iota})'B_{L(\iota)}(Z + h_{\iota})]^{2}} \right) \\
		 & \geq \mathbb{E} \left[\frac{tr(A_{L(\iota)}) - 2\phi_{max}(A_{L(\iota)})}{(Z + h_{\iota})'B_{L(\iota)}(Z + h_{\iota})} \right],
	\end{aligned}
\end{equation}
The inequality in \eqref{eq:risk_bound3} comes from the fact that
\begin{equation*}
	\frac{x'A_{L(\iota)}x}{x'x} \leq \max_{x} \frac{x'A_{L(\iota)}x}{x'x} = \phi_{\max}(A_{L(\iota)}),
\end{equation*}
which gives
\begin{equation*}
	(Z + h_{\iota})'\tilde{B}_{L(\iota)}'A_{L(\iota)}\tilde{B}_{L(\iota)}(Z + h_{\iota}) \leq (Z + h_{\iota})'B_{L(\iota)}(Z + h_{\iota}) \phi_{\max}(A_{L(\iota)}).
\end{equation*}

Combining \eqref{eq:risk_bound3} and \eqref{eq:risk_bound0}, we can show that
\begin{equation} \label{eq:risk_bound4}
	\begin{aligned}
		\mathbb{E}[\psi'W\psi] & \leq tr(W\Omega) - \tau\sum_{\iota = 2}^{2^{p}} \mathbb{E} \left[ \frac{
	2(tr(A_{L(\iota)}) - 2\phi_{\max}(A_{L(\iota)})) - \tau}{(Z + h_{\iota})'B_{L(\iota)}(Z + h_{\iota})} \right] \mathbb{P}(\tilde{\mu}_{\iota} > 0,\,\tilde{\mu}_{-\iota} \leq 0) \\
	& = tr(W\Omega) - \tau\sum_{\iota = 2}^{2^{p}} \mathbb{E} \left[ \frac{
	2(tr(A_{L(\iota)}) - 2\phi_{\max}(A_{L(\iota)})) - \tau}{\xi_{L(\iota)}} \right] \mathbb{P}(\tilde{\mu}_{\iota} > 0,\,\tilde{\mu}_{-\iota} \leq 0)
	\end{aligned}
\end{equation}
In order for the shrinkage estimator to have lower asymptotic risk than the unrestricted estimator, given $\tau > 0$, we require 
\begin{equation} \label{eq:tau_bound}
	\begin{aligned}
		& \sum_{\iota = 2}^{2^{p}}  
		\left[2(tr(A_{L(\iota)}) - 2\phi_{\max}(A_{L(\iota)})) - \tau \right] \mathbb{E}[\xi_{L(\iota)}^{-1}] \mathbb{P}(\tilde{\mu}_{\iota} > 0,\,\tilde{\mu}_{-\iota} \leq 0) \geq 0 \\
		\sum_{\iota = 2}^{2^{p}} 
		2(tr(A_{L(\iota)}) & - 2\phi_{\max}(A_{L(\iota)})) \mathbb{E}[\xi_{L(\iota)}^{-1}] \mathbb{P}(\tilde{\mu} > 0,\,\tilde{\mu}_{-\iota} \leq 0) - \tau \sum_{\iota = 2}^{2^{p}} \mathbb{E}[\xi_{L(\iota)}^{-1}] \mathbb{P}(\tilde{\mu}_{\iota} > 0,\,\tilde{\mu}_{-\iota} \leq 0) \geq 0 \\
		& \quad \quad \quad \quad \quad 0 < \tau \leq \sum_{\iota = 2}^{2^{p}} 2(tr(A_{L(\iota)}) - 2\phi_{\max}(A_{L(\iota)})) \gamma_{\iota},
	\end{aligned}
\end{equation}
where 
\begin{equation*}
	\gamma_{\iota} \equiv \frac{\mathbb{E}[\xi_{L(\iota)}^{-1}]\mathbb{P}(\tilde{\mu}_{\iota} > 0,\,\tilde{\mu}_{-\iota} \leq 0)}{\sum_{\iota = 2}^{2^{p}}\mathbb{E}[\xi_{L(\iota)}^{-1}]\mathbb{P}(\tilde{\mu}_{\iota} > 0,\,\tilde{\mu}_{-\iota} \leq 0)}.
\end{equation*}
Given the condition in \eqref{eq:tau_bound} and $h < \infty$, the risk in \eqref{eq:risk_bound4} is strictly less than $tr(W\Omega)$, which establishes \eqref{eq:risk_opt_bound}. \qed

\section{Mixtures of equality and inequality constraints} 

\renewcommand{\theequation}{B.\arabic{equation}}
\setcounter{equation}{0}

Some estimation problems involve a combination of equality and inequality constraints, e.g. estimating a parameter vector which represents probabilities, which have to be greater than zero and sum up to one. It turns out that it is straightforward to incorporate equality constraints into our analysis. To be specific, assume that there are $q$ equality constraints and $p-q$ inequality constraints. Since the distribution of the unconstrained estimator is unaffected by the composition of constraints, the main object of interest is the distribution of the constrained estimator, which determines the form of the shrinkage parameter.  

Following the intuition from Section \ref{sec:distribution}, the asymptotic distribution of the restricted estimator is given by
\begin{equation} \label{eq:asy_distr_mixed}
	n^{1/2}(\tilde{\theta}_{n} - \theta_{0}) \rightarrow_{d} Z - \sum_{\iota=1}^{2^{p-q}} P_{L(\iota)}(Z + h_{\iota})\mathbb{P}(\tilde{\mu}_{\iota} > 0,\,\tilde{\mu}_{-\iota} < 0).
\end{equation}
When none of the inequality constraints bind, $\iota = 1$, then the distribution in \eqref{eq:asy_distr_mixed} collapses to $Z - P_{L(1)}(Z+h_1)$, which is simply the distribution of the restricted estimator under $q$ equality constraints.

By analogy, the optimal level of shrinkage is
\begin{equation} 
	\tau^{*} = \sum_{\iota = 1}^{2^{p-q}} \left(tr(A_{L(\iota)}) - 2\phi_{\max}(A_{L(\iota)})\right) \gamma_{\iota},
\end{equation}
where
\begin{equation*}
	\gamma_{\iota} \equiv \frac{\mathbb{E}[\xi_{L(\iota)}^{-1}]\mathbb{P}(\tilde{\mu}_{\iota} > 0,\,\tilde{\mu}_{-\iota} \leq 0)}{\sum_{\iota = 1}^{2^{p-q}}\mathbb{E}[\xi_{L(\iota)}^{-1}]\mathbb{P}(\tilde{\mu}_{\iota} > 0,\,\tilde{\mu}_{-\iota} \leq 0)}.	
\end{equation*}
The presence of equality constraints makes it easier to satisfy condition \eqref{eq:tau_bounds}. For example, in case when $W = \Omega^{-1}$, if $q > 2$, then $\tau > 0$. The presence of equality constraints provides additional ex ante information about the shrinkage direction. However, the exact shrinkage direction will still depend on additional binding constraints.

\section{Example of a Linear model with sign restrictions} \label{app:lin_model}

\renewcommand{\theequation}{C.\arabic{equation}}
\setcounter{equation}{0}
\renewcommand{\thefigure}{C.\arabic{figure}}
\setcounter{figure}{0}

To illustrate the idea, we will derive the asymptotic distribution of the restricted estimator for the linear model with sign restrictions by solving an asymptotically equivalent problem.

Consider a simple linear model with sign restrictions. Let $y_{i} = x_{i}'\theta + \varepsilon_{i}$, where $\{(y_{i},\,x_{i},\,\varepsilon_{i})\}_{i=1}^{n}$ are iid. For simplicity assume that we only have two parameters, $\theta \in \mathbb{R}^{2}$, and we want to shrink towards all coefficients being non-negative, $\theta_{j} \geq 0$ for $j = 1,\,2$. In this example $\Theta_{0} = \{\theta \in {\Theta}: \theta \geq 0\}$, $r(\theta) = \theta$, $R = \mathcal{I}_{2}$, and the local to zero assumption becomes $\theta_{0} = cn^{-1/2}$.

The unconstrained Least Squares estimator is 
\begin{equation*}
	\hat{\theta}_{n} = (X'X)^{-1}(X'Y),
\end{equation*}
where $X$ is a $n \times 2$ matrix of stacked regressors and $Y$ is a $n \times 1$ output vector. The constrained estimator solves the following problem
\begin{equation} \label{eq:lin_constr_problem}
	\min_{\theta \in \Theta} \frac{1}{2}\Vert Y - X\theta \Vert^{2} \quad \text{s.t.} \quad \theta \geq 0.
\end{equation}

\begin{figure}[!t]
	\centering
	\begin{subfigure}{0.45\linewidth} 
    \begin{tikzpicture}[scale=1.2]
		\draw [thick, <->] (-1.5,2) -- (-1.5,-1.5) -- (2,-1.5);
		\node [below right] at (2,-1.5) {\footnotesize $\theta_{1}$};
		\node [left] at (-1.5,2) {\footnotesize $\theta_{2}$};
		\foreach \X in {0.5,0.7,0.9,1.1}
		{\draw[dashed, color=blue] plot[smooth cycle,tension=1,rotate=130] coordinates {(0,\X) (\X/2,0) (0,-\X) (-\X/2,0)};}
		\draw[fill] (0,0) circle [radius=0.025];
		\node [below right] at (0,0) {\footnotesize $\hat{\theta}_{n} = \tilde{\theta}_{n}$};
	\end{tikzpicture}
	\caption{None of the constraints bind.}
	\label{subfig:gr_1}
	\end{subfigure}
	\begin{subfigure}{0.45\linewidth} 
    \begin{tikzpicture}[scale=1.2]
		\draw [thick, <->] (1,2) -- (1,-1.5) -- (4,-1.5);
		\node [below right] at (4,-1.5) {\footnotesize $\theta_{1}$};
		\node [left] at (1,2) {\footnotesize $\theta_{2}$};
		\foreach \X in {0.5,0.7,0.9,1.1}
		{\draw[dashed, color=blue] plot[smooth cycle,tension=1,rotate=130] coordinates {(0,\X) (\X/2,0) (0,-\X) (-\X/2,0)};}
		\draw[fill] (0,0) circle [radius=0.025];
		\draw[->, thick] (0,0) node [below right] {\footnotesize $\hat{\theta}_{n}$} -- (1,0.8) node [below right] {\footnotesize $\tilde{\theta}_{n}$};
	\end{tikzpicture}
	\caption{The first constraint binds, $\theta_{1} = 0$.}
	\label{subfig:gr_2}
	\end{subfigure}
	\begin{subfigure}{0.45\linewidth} 
	\begin{tikzpicture}[scale=1.2]
		\draw [thick, <->] (-2.3,3) -- (-2.3,1) -- (1.3,1);
		\node [below right] at (1.3,1) {\footnotesize $\theta_{1}$};
		\node [left] at (-2.3,3) {\footnotesize $\theta_{2}$};
		\foreach \X in {0.5,0.7,0.9,1.1}
		{\draw[dashed, color=blue] plot[smooth cycle,tension=1,rotate=130] coordinates {(0,\X) (\X/2,0) (0,-\X) (-\X/2,0)};}
		\draw[fill] (0,0) circle [radius=0.025];
		\draw[->, thick] (0,0) node [below right] {\footnotesize $\hat{\theta}_{n}$} -- (0.85,1) node [above left] {\footnotesize $\tilde{\theta}_{n}$};
	\end{tikzpicture}
	\caption{The second constraint binds, $\theta_{2} = 0$.}
	\label{subfig:gr_3}
	\end{subfigure}
	\begin{subfigure}{0.45\linewidth} 
    \begin{tikzpicture}[scale=1.2]
		\draw [thick, <->] (1,3) -- (1,1) -- (4,1);
		\node [below right] at (4,1) {\footnotesize $\theta_{1}$};
		\node [left] at (1,3) {\footnotesize $\theta_{2}$};
		\foreach \X in {0.5,0.7,0.9,1.1}
		{\draw[dashed, color=blue] plot[smooth cycle,tension=1,rotate=130] coordinates {(0,\X) (\X/2,0) (0,-\X) (-\X/2,0)};}
		\draw[fill] (0,0) circle [radius=0.025];
		\draw[->, thick] (0,0) node [below right] {\footnotesize $\hat{\theta}_{n}$} -- (1,1) node [below right] {\footnotesize $\tilde{\theta}_{n}$};
	\end{tikzpicture}
	\caption{Both constraints bind.}
	\label{subfig:gr_4}
	\end{subfigure}
	\vspace{1em}
	\mycaption{Geometry of the restricted estimator}{This figure shows the behavior of the restricted estimator depending on the position of the true parameter value in the two-dimensional case, $\theta_1 \geq 0$ and $\theta_2 \geq 0$. The blue dashed lines are LS objective contour sets.}
	\label{fig:geom_restricted}
\end{figure}

A solution to \eqref{eq:lin_constr_problem} depends on which constraints actually hold, which is in turn driven by the location of the true parameter value $\theta_{0}$ relative to the restricted space $\Theta_{0}$. Figure \ref{fig:geom_restricted} illustrates the main idea. When $\theta_{0} \in \Theta_{0}$, meaning none of the constraints bind, the restricted estimator coincides with the unrestricted one, $\hat{\theta}_{n} = \tilde{\theta}_{n}$ (see Figure \ref{subfig:gr_1}). However, if $\theta_{0} \not \in \Theta_{0}$, the restricted estimator becomes a projection of the unrestricted estimator on the closest boundary. In Figure \ref{subfig:gr_2}, $\theta_{0}$ is close to the boundary where $\theta_1 = 0$ and $\theta_2 > 0$, hence, $\tilde{\theta}_{n}$ is a projection of $\hat{\theta}_{n}$ on the half-space $\{\theta:\theta_1 = 0,\,\theta_2 > 0\}$. Figure \ref{subfig:gr_3} demonstrates the reciprocal case where $\tilde{\theta}_{n}$ is a projection on $\{\theta:\theta_1 > 0,\,\theta_2 = 0\}$. Finally, in Figure \ref{subfig:gr_4}, $\tilde{\theta}_{n}$ is a projection on $\{\theta:\theta_1 = 0,\,\theta_2 = 0\}$.

Let 
\begin{align*}
	& \mathcal{J}_{n} = n^{-1}X'X, \quad \mathcal{J} = \mathbb{E}[x_{i}x_{i}'], \quad Z_{n} = n^{1/2}(X'X)^{-1}X'\varepsilon, \\ 
	& \quad Z = \mathcal{J}^{-1}G, \quad G \sim \mathcal{N}(0,\,\mathcal{V}), \quad \mathcal{V} = \mathbb{E}[x_{i}x_{i}'\varepsilon_{i}^{2}]. 
\end{align*}

The quadratic approximation of the objective function takes the form
\begin{equation*}
	q_{n}(\lambda) = \frac{1}{2n} (\lambda - Z_{n})'(X'X)(\lambda - Z_{n}).
\end{equation*}
The vector of Kuhn-Tucker multipliers is 
\begin{equation*}
	\tilde{\mu}_{n} = -(n^{-1}X'X)^{-1}(Z_{n} + c) \hspace{0.1cm} \stackrel{d}{\rightarrow} \hspace{0.1cm}\tilde{\mu} = -\mathcal{J}^{-1}(Z + c). 
\end{equation*}

All constraints are satisfied as equalities if $\tilde{\mu}_{n,j} > 0$ for $j = 1,\,2$. In this case the restricted estimator is equal to zero, $\tilde{\theta}_{n}=0$, and $\tilde{\lambda}_{n} = -c$. All constraints are satisfied as strict inequalities if $\tilde{\mu}_{n,j} \leq 0$ for $j = 1,\,2$. Then the restricted estimator equals to the unrestricted one, $\tilde{\theta}_{n} = \hat{\theta}_{n}$, and $\tilde{\lambda}_{n} = Z$. The first constraint binds if $\tilde{\mu}_{n,1} \leq 0$ and $\tilde{\mu}_{n,2} > 0$, which implies that $\tilde{\lambda}_{1} = -c_{1}$ and $\tilde{\lambda}_{2} = Z_{2} - \mathcal{J}^{-1}_{21}\mathcal{J}_{11}(Z_{1} + c_{1})$.\footnote{The Hessian matrix here is just $2 \times 2$. Therefore, the sub-matrices in the expression for $\tilde{\lambda}_2$ are just the corresponding elements of $\mathcal{J}$.} We get a similar result for the case where the second constraint binds. Therefore, according to Lemma \ref{lemma:distr_re}, the resulting asymptotic distribution of the restricted estimator takes the form
\begin{equation*} \label{eq:asy_distr_exmp}
	\begin{aligned}
		n^{1/2} 
		\begin{pmatrix}
			\tilde{\theta}_{n,1} - \theta_{0,1} \\ \tilde{\theta}_{n,2} - \theta_{0,2}
		\end{pmatrix}
		\hspace{0.1cm} \stackrel{d}{\rightarrow} \hspace{0.1cm} &
		\begin{pmatrix}
			Z_{1} \\ Z_{2}
		\end{pmatrix}
		\mathds{1}\{ \tilde{\mu}_{1} \leq 0,\, \tilde{\mu}_{2} \leq 0 \} \\
		+ & 
		\begin{pmatrix}
			-c_{1} \\ Z_{2} - \mathcal{J}^{-1}_{21}\mathcal{J}_{11}(Z_{1} + c_{1})		
		\end{pmatrix}
		\mathds{1}\{ \tilde{\mu}_{1} > 0,\, \tilde{\mu}_{2} \leq 0 \} \\
		+ & 
		\begin{pmatrix}
			Z_{2} - \mathcal{J}^{-1}_{12}\mathcal{J}_{22}(Z_{2} + c_{2})	 \\ -c_{2}	
		\end{pmatrix}
		\mathds{1}\{ \tilde{\mu}_{1} \leq 0,\, \tilde{\mu}_{2} > 0 \} \\
		+ & 
		\begin{pmatrix}
			-c_{1} \\ -c_{2}
		\end{pmatrix}
		\mathds{1}\{ \tilde{\mu}_{1} > 0,\, \tilde{\mu}_{2} > 0 \} .
	\end{aligned}
\end{equation*}

\section{Empirical Bayes Estimator} \label{app:eb}

In matrix form the linear model from Section \ref{sec:mc} is $Y = X\theta + \varepsilon$. Hence, the likelihood density is
\begin{equation*}
	p(Y|\theta) = (2\pi)^{-n/2}\exp\left\{-\frac{1}{2}(Y - X\theta)'(Y - X\theta)\right\}.
\end{equation*}
We assume the prior $\theta|\nu \sim \mathcal{N}(0,\,1/\nu)\mathds{1}\{\theta \geq 0\}$, which density is
\begin{align*}
	p(\theta|\nu) & = \Pi_{j=1}^{m}\left(\frac{2\pi}{\nu}\right)^{-1/2}\mathbb{P}_{\theta|\nu}^{-1}(\theta_{j} \geq 0)\exp\left\{-\frac{\nu}{2}\theta_{j}^{2}\right\} \\
	& = \left(\frac{2\pi}{\nu}\right)^{-m/2}\Phi(0)^{-m}\exp\left\{-\frac{\nu}{2}\theta'\theta\right\},
\end{align*}
where $\Phi(\cdot)$ is the cdf of the standard normal distribution.

By Bayes' rule, the posterior distribution is
\begin{align*}
	p(\theta|Y,\,\nu) & \propto p(Y|\theta) p(\theta|\nu) \\
	& \propto \exp\left\{-\frac{1}{2}(\theta - \bar{\theta})'\bar{V}_{\theta}^{-1}(\theta - \bar{\theta})) \right\}\mathds{1}\{\theta \geq 0\},
\end{align*}
where 
\begin{align*}
	\bar{\theta} & = (X'X + \nu\mathcal{I}_{m})^{-1}X'Y \\
	\bar{V}_{\theta} & = (X'X + \nu\mathcal{I}_{m})^{-1}.
\end{align*}
Thus, the posterior has also a truncated normal form, $\theta|Y,\,\nu \sim \mathcal{N}(\bar{\theta},\,\bar{V}_{\theta})\mathds{1}\{\theta \geq 0\}$.

The marginal likelihood is
\begin{equation*}
	p(Y|\nu) = \int p(Y|\theta) p(\theta|\nu) d\theta.
\end{equation*}
We can use the Bayes' rule again to get an explicit expression for the marginal likelihood density
\begin{align*}
	p(Y|\nu) & = \frac{p(Y|\theta) p(\theta|\nu)}{p(\theta|Y,\,\nu)} \\
	& = (2\pi)^{-n/2}\nu^{m/2}\Phi(0)^{-m}|\bar{V}_{\theta}|^{1/2}D \exp\left\{-\frac{1}{2}[Y'Y - Y'X(X'X + \nu\mathcal{I}_{m})^{-1}X'Y] \right\},
\end{align*}
where $D = \mathbb{P}_{\theta|Y,\,\nu}(\theta \geq 0)$ is a normalizing constant for the the posterior. Note that $D$ depends on $\nu$.

We select $\nu$ by maximizing the marginal likelihood density:
\begin{equation*}
	\hat{\nu} = \argmax_{\nu \in \mathbb{R}_{+}}p(Y|\nu).
\end{equation*}
Then the Empirical Bayes estimator is simply the mean of the posterior distribution given $\hat{\nu}$, $p(\theta|Y,\,\hat{\nu})$.

\section{Empirical Application: Implementation Details} \label{app:ea_details}

\renewcommand{\theequation}{E.\arabic{equation}}
\setcounter{equation}{0}

The set-up is close to \cite{kasy2018}. Let $Q,\,P$, and $Y$ denote the quantity (of gasoline in our application) demanded by a consumer, the price paid, and the consumer's income. Assume that we observe data $\{Q_{i},\,P_{i},\,Y_{i}\}_{i=1}^{n}$ on $n$ randomly sampled consumers. We assume that the variables are related as
\begin{equation*}
	Q = g(P,\,Y) + U, 
\end{equation*}
where $g$ is an unknown demand function, and $U$ satisfies $\mathbb{E}[U|P=p,\,Y=y] = 0$ for all $p$ and $y$. The latter assumption on the unobserved shock assumes that prices and incomes are statistically independent of unobserved preference heterogeneity across consumers. That is, we ignore endogeneity concerns for the ease of exposition. 

Our goal is to estimate the price and income elasticities of $g(p,\,y)$, $\beta_{j}^{p}$ and $\beta_{j}^{y}$, respectively, at different price levels $p_{1},\dots,\,p_{J}$ and a given income level $y$,
\begin{equation*}
	\beta_{j}^{p} = \frac{\partial \log g(p_{j},\,y)}{\partial \log p}, \quad \beta_{j}^{y} = \frac{\partial \log g(p_{j},\,y)}{\partial \log y}.
\end{equation*}
We can obtain the unrestricted elasticities estimates for a price-income pair $(p_{j},\,y)$ using a local linear regression (LLR),
\begin{multline} \label{eq:llr}
	\left(\hat{\alpha}_{j},\,\hat{\beta}_{j}^{p},\,\hat{\beta}_{j}^{y}\right) = \argmin_{a,\,b^{p},\,b^{y}} \sum_{i=1}^{n} \left(Q_{i} - a - b^{p} (\log P_{i} - \log p_{j}) - b^{y} (\log Y_{i} - \log y) \right)^2 \times \\  
	K_{h}\left(\log P_{i} - \log p_{j},\, \log Y_{i} - \log y\right),
\end{multline}
where $\hat{\alpha}_{j} = \hat{g}(p_{j},\,y)$, and $K_{h}$ is a kernel function with bandwidth $h$ (we use the Epanechnikov kernel). We use nonparametric bootstrap to estimate the joint variance $V$ of $\hat{\beta} \equiv \left(\hat{\beta}_{j}^{p},\,\hat{\beta}_{j}^{y}\right)_{j=1}^{J}$ across all $j$. As in \cite{kasy2018}, the variance of $\hat{\alpha}_{j}$ is negligible compared to $V$ in our application.

Slutsky condition is an inequality constraint on the demand function ensuring that the compensated own-price elasticities are negative,
\begin{equation*}
	\frac{\partial g(p_{j},\,y)}{\partial p} + \frac{\partial g(p_{j},\,y)}{\partial y}g(p_{j},\,y) \leq 0, \quad j = 1,\dots,\,J.
\end{equation*}
Rewritten in terms of elasticities, it gives us the desired theoretical restriction 
\begin{equation} \label{eq:slutsky_restriction}
	\beta_{j}^{p} + \beta_{j}^{y} g(p_{j},\,y) \frac{p_{j}}{y} \leq 0, \quad j = 1,\dots,\,J.	
\end{equation}
The restricted estimator $\tilde{\beta} \equiv \left(\tilde{\beta}_{j}^{p},\,\tilde{\beta}_{j}^{y}\right)_{j=1}^{J}$ solves \eqref{eq:llr} under the condition \eqref{eq:slutsky_restriction} for all $j$.

The ICSE takes the weighted average form $\hat{\beta}^{\ast} = \hat{w}\hat{\beta} + (1 - \hat{w})\tilde{\beta}$ where 
\begin{equation*}
	\hat{w} = \left(1 - \frac{\hat{\tau}^{\ast}}{n(\hat{\beta} - \tilde{\beta})'(\hat{\beta} - \tilde{\beta})}\right)_{+},
\end{equation*}
and $\hat{\tau}^{\ast}$ is given by \eqref{eq:feasible_icse}. Note that the feasible ICSE shrinks $\hat{\beta}$ to $\tilde{\beta}$ jointly across all $j$.

\end{appendices}

\end{document}